\documentclass[%
reprint,
 amsmath,amssymb,
 aps,
]{revtex4-1}

\usepackage{graphicx}
\usepackage{array}
\usepackage{dcolumn}
\usepackage{bm}
\usepackage{listings}
\usepackage{scalefnt}
\usepackage{xcolor}
\usepackage{booktabs}

\def\bra#1{\mathinner{\langle{#1}|}}
\def\ket#1{\mathinner{|{#1}\rangle}}
\newcommand{\braket}[2]{\langle #1|#2\rangle}

\newcommand{\ra}[1]{\renewcommand{\arraystretch}{#1}}

\begin{document}

\title{Boson Sampling for Molecular Vibronic Spectra}

\author{Joonsuk Huh}
\email{Email: huh@fas.harvard.edu}
\affiliation{Department of Chemistry and Chemical Biology, Harvard University, Cambridge, Massachusetts 02138, United States}
\author{Gian Giacomo Guerreschi}
\affiliation{Department of Chemistry and Chemical Biology, Harvard University, Cambridge, Massachusetts 02138, United States}
\author{Borja Peropadre}
\affiliation{Department of Chemistry and Chemical Biology, Harvard University, Cambridge, Massachusetts 02138, United States}
\author{Jarrod R. McClean}
\affiliation{Department of Chemistry and Chemical Biology, Harvard University, Cambridge, Massachusetts 02138, United States}
\author{Al\'an Aspuru-Guzik}
\email{Email: aspuru@chemistry.harvard.edu}
\affiliation{Department of Chemistry and Chemical Biology, Harvard University, Cambridge, Massachusetts 02138, United States}
\date{\today}

\begin{abstract}
Quantum computers are expected to be more efficient in performing certain computations
than any classical machine. Unfortunately, the technological challenges associated with
building a full-scale quantum computer have not yet allowed the experimental verification
of such an expectation. Recently, boson sampling has emerged as a problem that is suspected
to be intractable on any classical computer, but efficiently implementable with a linear
quantum optical setup. Therefore, boson sampling may offer an experimentally realizable
challenge to the Extended Church-Turing thesis and this remarkable possibility motivated
much of the interest around boson sampling, at least in relation to
complexity-theoretic questions. In this work, we show that the successful development
of a boson sampling apparatus would not only answer such inquiries, but also yield
a practical tool for difficult molecular computations.  Specifically, we show that a boson
 sampling device with a modified input state can be used to generate molecular vibronic
spectra, including complicated effects such as Duschinsky rotations.  

\end{abstract}

\maketitle


\section{Introduction}
\begin{table*}[htb]
  \centering
  \ra{1.3}
  \begin{tabular}{@{}ccc@{}}
  \hline\\
     & Boson Sampling & Vibronic Transitions \\
     \hline\\
    
    Harmonic oscillators &
    \begin{minipage}{.4\textwidth}
      \includegraphics[width=0.5\linewidth]{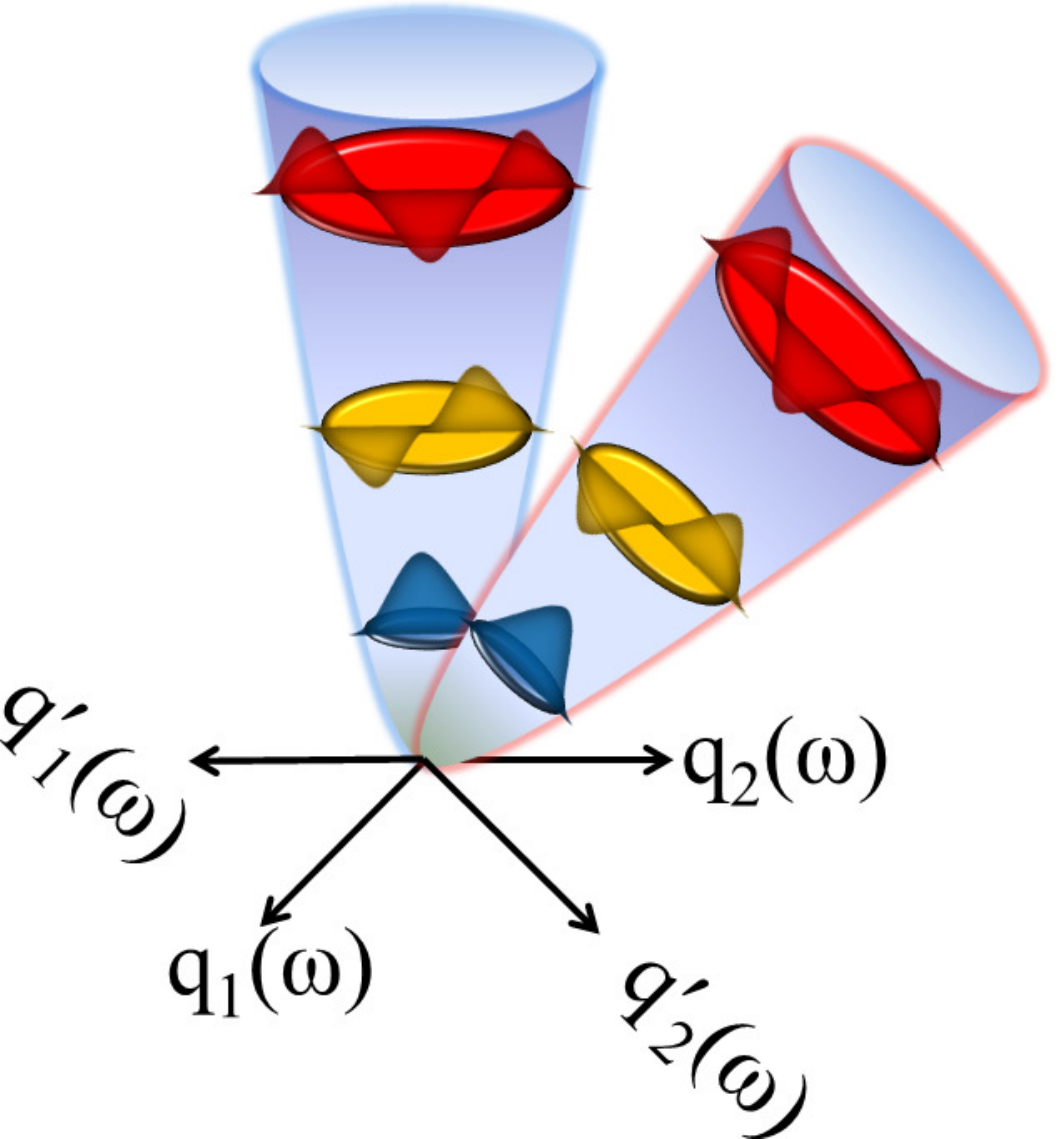}
    \end{minipage}
    
    & 
    \begin{minipage}{.4\textwidth}
      \includegraphics[width=0.7\linewidth]{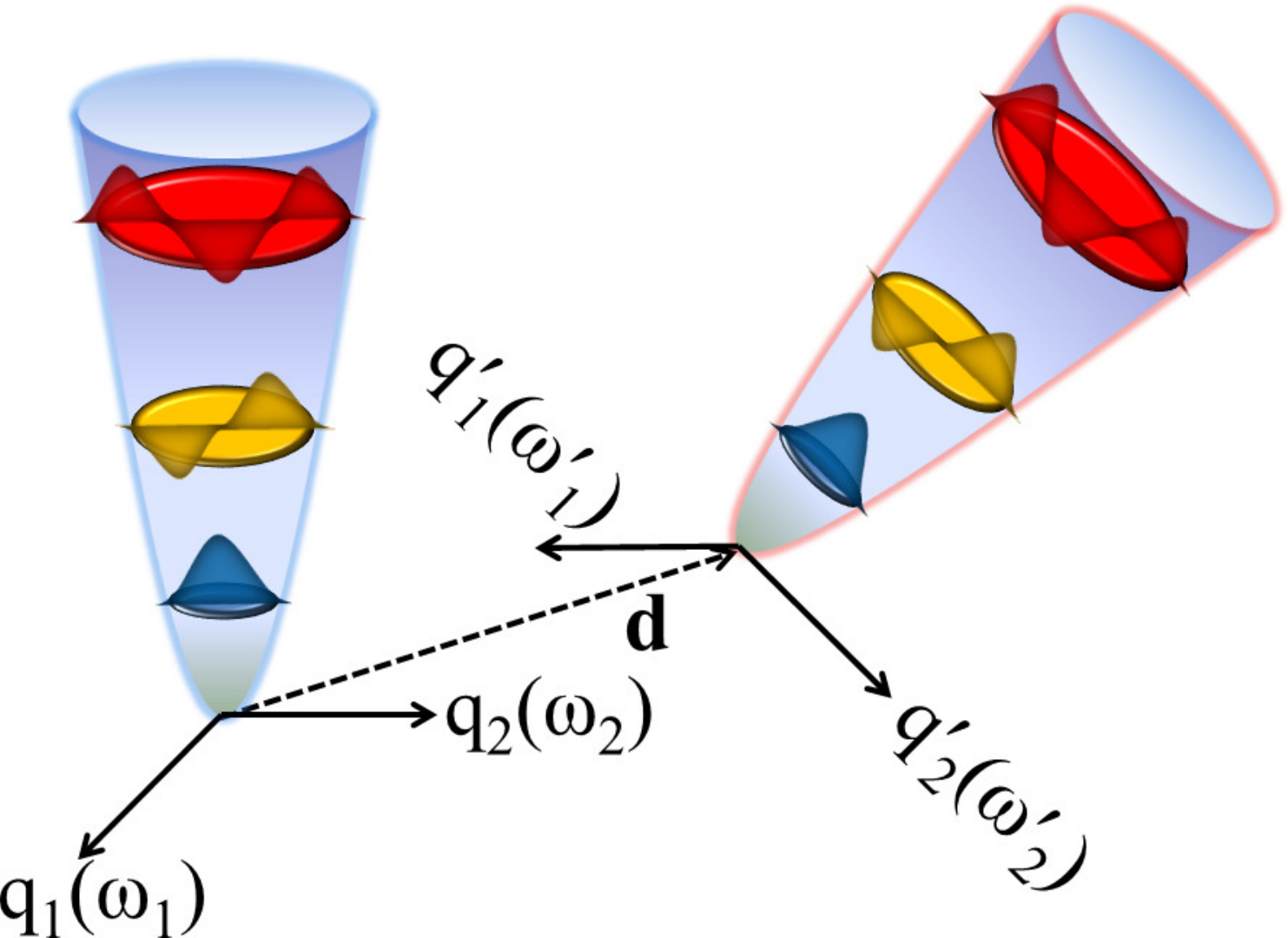}
    \end{minipage}
    \\ 
    Linear transform & $\mathbf{\hat{a}}^{'\dagger}=\mathbf{U}\mathbf{\hat{a}}^{\dagger}$& 
    $\mathbf{\hat{a}}^{'\dagger}=\frac{1}{2}\left(\mathbf{J}-(\mathbf{J}^{\mathrm{t}})^{-1}\right)\mathbf{\hat{a}}
    +\frac{1}{2}\left(\mathbf{J}+(\mathbf{J}^{\mathrm{t}})^{-1}\right)\mathbf{\hat{a}}^{\dagger}+\frac{1}{\sqrt{2}}\boldsymbol{\delta}$\\ 
    Unitary operators & Rotation & Displacement, Squeezing and Rotation \\ 
    Particle to simulate & Photon & Phonon\\ 
    Particle in simulator & Photon & Photon\\ 
    Outcome of simulation & $\vert$Permanent$\vert^{2}$ & Franck-Condon profile (spectrum)\\
    \hline
  \end{tabular}
	\caption{A comparison of boson sampling and the computation of vibrionic transitions. The quantum harmonic oscillators (QHOs) in the first
    row show the corresponding two-dimensional normal coordinates ($q_{k}$ and $q_{l}'$ for input and
    output states, respectively) and their respective harmonic frequencies ($\omega_{k}$ and $\omega_{l}'$). The
    two sets of QHOs in boson sampling are rotated with respect to each other such that the linear
    relation with the rotation matrix $\mathbf{U}$ of the boson creation operators are given in the
    second row. The two sets of QHOs in vibronic transitions are displaced, distorted (frequency changes)
    and rotated with respect to each other. $\mathbf{d}$ is a displacement vector of the quantum
    harmonic oscillators. The boson creation operator ($\mathbf{\hat{a}}^{'\dagger}$) of the output
    state is now given as a linear combination of the boson annihilation ($\mathbf{\hat{a}}$) and
    creation ($\mathbf{\hat{a}}^{\dagger}$) operators of the input state with the dimensionless
    displacement vector $\boldsymbol{\delta}$. A matrix $\mathbf{J}$ characterizes the rotation and
    squeezing operations during a vibronic transition.  Note that this picture applies only when
    $\mathbf{U}$ is a real matrix.}
	\label{tab:bsvib}
\end{table*}

Quantum mechanics allows the storage and manipulation of information in ways that are not
possible according to classical physics. At a glance, it appears evident that
the set of operations characterizing a quantum computer is strictly larger than the
operations possible in a classical hardware. This speculation is at the basis of quantum
speedups that have been achieved for oracular and search problems \cite{Deutsch1992,Grover1997}.
Particularly significant is the exponential speed up achieved for the prime factorization
of large numbers \cite{Shor1999}, a problem for which no efficient classical algorithm is
currently known.
Another attractive area for quantum computers is quantum simulation
\cite{Georgescu2014,Aspuru-Guzik2012,Bloch2012,Blatt2012,Lloyd1996,Aspuru-Guzik2005}
where it has recently been shown that the dynamics of chemical reactions \cite{Kassal2011}
as well as molecular electronic structure~\cite{Babbush2014} 
are attractive applications for quantum devices.
For all these instances, the realization of a quantum computer would challenge the Extended
Church-Turing thesis (ECT), which claims that a Turing machine can efficiently simulate any
physically realizable system, and even disprove it if prime factorization was finally demonstrated
to be not efficiently solvable on classical machines. 

\begin{figure}[htb]
\begin{center}
\includegraphics[width=\linewidth]{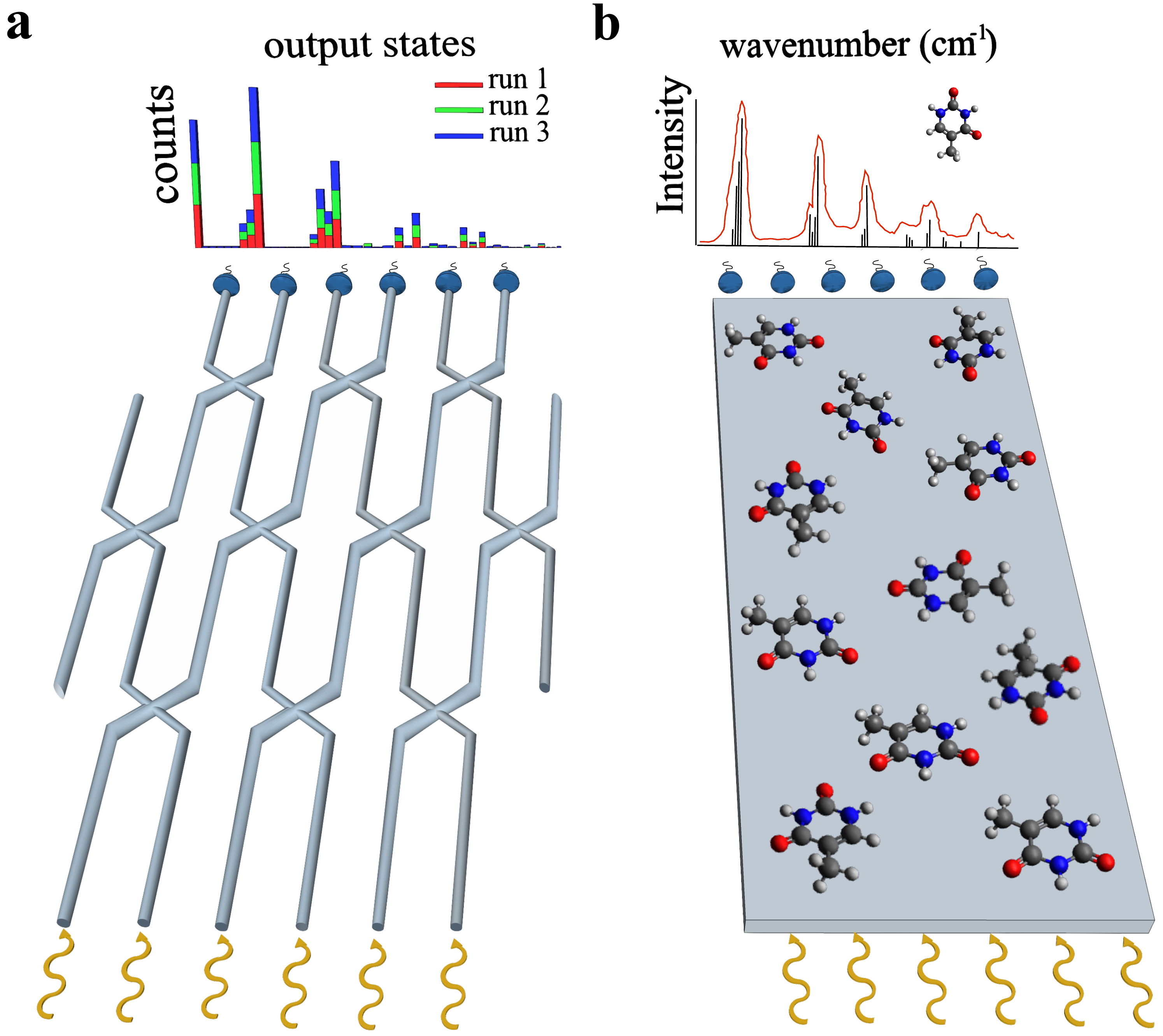}
\caption{Pictorial description of boson sampling and molecular vibronic spectroscopy. {\bf a}, Boson sampling
consists of sampling the output distribution of photons obtained from quantum interference inside a
linear quantum optical network. {\bf b}, Vibronic spectroscopy uses coherent light to electronically
excite an ensemble of identical molecules and measures the re-emitted (or scattered) radiation to
infer the vibrational spectrum of the molecule. We show in this work how the fundamental physical
process underlying {\bf b} is formally equivalent to situation {\bf a} together with a non-linear state
preparation step.
}\label{fig:pictorial}
\end{center}
\end{figure}

At the same time, the realization of a full-scale quantum computer is a very demanding  technological challenge, 
even if it is not forbidden by fundamental physics.
This fact motivated the search for intermediate quantum hardware that could efficiently solve
specific computational problems, believed to be intractable with classical machines, without
being capable of universal quantum computation.
Recently, Aaronson and Arkhipov found that sampling the distribution of photons at the
output of a linear photonic network is expected (modulo a few conjectures) to be
computationally inefficient for any classical computer since it would require the estimation
of lots of matrix permanents \cite{Aaronson2011}. On the contrary, this task is naturally
simulated by indistinguishable photons injected as input of a photonic network
(see the pictorial description of boson sampling in Fig.~\ref{fig:pictorial}{\bf a}). 
While several groups have already realized small-scale versions of boson sampling
\cite{Spring2013,Broome2013,Crespi2013,Tillmann2013}, to challenge the ECT one also has
to demonstrate the scalability of the experimental architecture \cite{Shchesnovich2014,Rohde2014}. 

While boson sampling is likely to play a major role in the debate around the ECT, it also
appears as a somewhat artificial problem in which we ask a classical computer to predict
the behaviour of a quantum machine (under certain working conditions) and then compare its
efficiency to the direct operation of the machine itself.  In this work, we present a
connection between boson sampling and the calculation of molecular vibronic
(vibrational and electronic) spectra related to molecular processes such as absorption, emission, photoelectron and
resonance Raman (see Table~\ref{tab:bsvib}) 
\cite{sharp:1964,doktorov:1977,malmqvist:1998,cjruhoff:2000,jankowiak:2007,santoro:2007b} . 
The proposed simulation scheme provides a second, chemically-relevant reason to realize boson sampling machines.
The calculation of Franck-Condon (FC) factors with Duschinsky
mode mixing ~\cite{duschinsky:1937} represent a computationally difficult problem and various
strategies have been developed to overcome the difficulties (see \emph{e.g.}
Refs.~\cite{cjruhoff:2000,jankowiak:2007,santoro:2007b}). 
We show that the quantum simulation, and hence the calculation of FC factors lying at the heart
of linear spectroscopy, can be efficiently performed on a boson sampling machine simply by
modifying the input state.  This connection provides a scientific and industrially relevant
problem with a physical and chemical meaning well separated from the simulation of
linear quantum optical networks. 
A complementary approach for the quantum simulation of molecular vibrations in quantum optics using a time domain approach was recently introduced by Laing \emph{et al.}~\cite{lopez2014}. 

This work is organized as follows.  First, we introduce the boson sampling problem and
the Duschinsky relation in terms of unitary operators, then we compare
the two physical problems and suggest how to obtain molecular vibronic spectra with a
boson sampling device. We also provide examples concerning the photoelectron spectra of
formic acid and thymine~\cite{jankowiak:2007}. Finally, we conclude with an outlook for
the use of boson sampling devices in chemistry and molecular physics.

\section{Boson Sampling vs. Molecular Vibronic Transitions}

\subsection{Boson Sampling and Vibrational Overlaps}
Boson sampling considers the input of $N$ photons into $M$ optical modes. 
This quantum space can be described through a Fock basis that takes notice
of the number of photons distributed in each mode: We denote such states by
$\ket{n_1, n_2, ..., n_M} = \ket{\mathbf{n}}$, where $n_j$ corresponds to the number of
photons in the $j$-th mode and we have the constraint $\sum_j n_j = N$. These photons
are sent through a linear optical network whose action is characterized
by the unitary operation $\hat{U}$. In this way, any input state
$\ket{\phi_{\text{in}}}$ is related to the corresponding output state $\ket{\phi_{\text{out}}}$
through the relation,

\begin{align}
\ket{\phi_{\text{out}}} = \hat{U}\ket{\phi_{\text{in}}} \, .
\end{align}

Given this setup, the problem is to compute the transition probability between input and
output states in the Fock basis expressed by the quantity, 
\begin{align}
P_{\mathbf{nm}} = |\bra{\mathbf{m}} \hat{U} \ket{\mathbf{n}}|^2 \, ,
\end{align}
where $\ket{\mathbf{n}}$ is the input state and $\ket{\mathbf{m}}$ the desired state in
output. To facilitate the notation, we introduce the states,
\begin{align}
\ket{\mathbf{n};\mathrm{in}}  &= \ket{\mathbf{n}}, \\
\ket{\mathbf{m};\mathrm{out}} &= \hat{U}^{\dagger} \ket{\mathbf{m}} \, ,
\end{align}
to explicitly label which states have passed through the boson sampling apparatus.
As the total number of
photons and the number of modes increase, the probability distribution of output states 
becomes hard to predict with classical computers, but it can be directly measured with linear
optics devices. The computation of this distribution can be performed in a number of ways,
but here, to facilitate the forthcoming connections, we follow a construction that is similar
to those used in vibronic spectroscopy.

The basis set expansion of the input state ($\vert \mathbf{n};\mathrm{in}\rangle$) in the output Fock
states ($\vert \mathbf{m};\mathrm{out}\rangle$) is given by, 

\begin{align}
\vert \mathbf{n};\mathrm{in}\rangle=\sum_{\mathbf{m}\in \mathbb{S}_{N}}\langle\mathbf{m};\mathrm{out}\vert \mathbf{n};\mathrm{in}\rangle\vert \mathbf{m};\mathrm{out}\rangle
\end{align}
where $\mathbb{S}_{N}$ is the set of all possible $\mathbf{m}$ having the total number of photons $N =\sum_{k}m_{k}$.
Coherent states can be expanded in terms of Fock states and the overlap integrals
$\langle\mathbf{m};\mathrm{out}\vert \mathbf{n};\mathrm{in}\rangle$ can be obtained from the overlap integral of
two coherent states ($\langle\boldsymbol{\gamma};\mathrm{out}\vert\boldsymbol{\alpha};\mathrm{in}\rangle$), which are rotated with respect to each other in the phase space~\cite{doktorov:1977}. 
$\boldsymbol{\gamma}$ and $\boldsymbol{\alpha}$ are $M$-dimensional complex (column) vectors defining the
coherent states $\ket{\boldsymbol{\gamma};\mathrm{out}}=\hat{R}_{\mathbf{U}}^{\dagger}\ket{\boldsymbol{\gamma};\mathrm{in}}$ and $\ket{\boldsymbol{\alpha};\mathrm{in}}$. 
The unitary rotation operator $\hat{R}_{\mathbf{U}}$ acts on a column vector of creation operators
$\hat{\mathbf{a}}^{\dagger}=(\hat{a}^{\dagger}_{1},\ldots,\hat{a}^{\dagger}_{M})^{\mathrm{t}}$ and to produce a vector of rotated creation operators $\hat{\mathbf{a}}^{'\dagger}=(\hat{a}^{'\dagger}_{1},\ldots,\hat{a}^{'\dagger}_{M})^{\mathrm{t}}$
defined as, 

\begin{align}
\hat{\mathbf{a}}^{'\dagger} = \hat{R}_{\mathbf{U}}^{\dagger}\hat{\mathbf{a}}^{\dagger}\hat{R}_{\mathbf{U}}
							= \mathbf{U}\hat{\mathbf{a}}^{\dagger} .
\label{eq:operatorrotation} 
\end{align}
where we use a shorthand notation~\cite{Ma1990} for the operator action on the column vector of the boson creation operators, \emph{i.e.}
\begin{align}
\hat{A}\hat{\mathbf{a}}^{\dagger}\hat{B}=(\hat{A}\hat{a}^{\dagger}_{1}\hat{B},\ldots,\hat{A}\hat{a}^{\dagger}_{M}\hat{B})^{\mathrm{t}} \,  .
\end{align}
It is straightforward to verify that the bosonic commutation relations $[\hat{a}_{k},\hat{a}_{l}^{\dagger}]=\delta_{kl}$
are satisfied also for the rotated operators, \emph{i.e.} $[\hat{a}_{k}',\hat{a}_{l}^{'\dagger}]=\delta_{kl}$. 



One can obtain the transition amplitude $\langle\mathbf{m};\mathrm{out}\vert \mathbf{n};\mathrm{in}\rangle$
in two distinct ways: One way involves taking partial derivatives of 
$\mathrm{exp}(\tfrac{1}{2}(\vert\boldsymbol{\gamma}\vert^{2}+\vert\boldsymbol{\alpha}\vert^{2}))\langle\boldsymbol{\gamma};\mathrm{out}\vert\boldsymbol{\alpha};\mathrm{in}\rangle$~\cite{doktorov:1977}:

\begin{align}
\langle\mathbf{m};\mathrm{out}\vert \mathbf{n};\mathrm{in}\rangle
	&=\left(\prod_{k=1}^{M}\frac{\partial_{\alpha_{k}}^{n_{k}}\partial_{\gamma_{k}^{*}}^{m_{k}}}{\sqrt{n_{k}! m_{k}!}}\right)
\exp\left(\boldsymbol{\gamma}^{\dagger}\mathbf{U}^{*}\boldsymbol{\alpha}\right) \Biggr\rvert_{\boldsymbol{\alpha},\boldsymbol{\gamma}^{*}=\mathbf{0}} \, ,
\label{eq:partial} 
\end{align}
where $\partial_{a}^{b}=\tfrac{\partial^{b}}{\partial a^{b}}$ , and the second approach involves computing matrix
permanents (Per) of submatrices of $\mathbf{U}$~\cite{Scheel2004,Aaronson2011} as described by
\begin{align}
\langle\mathbf{m};\mathrm{out}\vert \mathbf{n};\mathrm{in}\rangle
	&=\left(\prod_{k=1}^{M}\sqrt{n_{k}! m_{k}!}\right)^{-1}\left(\mathrm{Per}\left([\mathbf{U}]_{\mathbf{n},\mathbf{m}}\right)\right)^{*} \, , \label{eq:permanent}
\end{align}
with $[\mathbf{U}]_{\mathbf{n},\mathbf{m}}$ being a $N\times N$ submatrix of $\mathbf{U}$ obtained by
repeating the $k$-th column of $\mathbf{U}$ $n_{k}$ times and copying the $l$-th row of the column of
the resulting matrix $m_{l}$ times~\cite{Aaronson2011}. 
The transition probability measured in boson sampling setups, which is proportional to
$\vert\langle\mathbf{m};\mathrm{out}\vert\mathbf{n};\mathrm{in}\rangle\vert^{2}$,
can be considered a calculation of the permanent of the matrix described above through
the relation~\cite{Scheel2004,Aaronson2011},
\begin{align}
\left\vert\mathrm{Per}\left([\mathbf{U}]_{\mathbf{n},\mathbf{m}}\right)\right\vert^2
	=\left(\prod_{k=1}^{M} n_{k}! m_{k}!\right)
		\left\vert\langle\mathbf{m};\mathrm{out}\vert \mathbf{n};\mathrm{in}\rangle\right\vert^2 \, .
\end{align}

From this, it is clear that the output of boson sampling devices can be related to both the permanent
of submatrices of $\mathbf{U}$ as well as partial to the derivatives of 
$\exp\left(\boldsymbol{\gamma}^{\dagger}\mathbf{U}^{*}\boldsymbol{\alpha}\right)$.  The calculation of matrix permanents is a
computationally hard problem for many classes of matrices belonging to the complexity class \#P~\cite{Aaronson2011}, and this
suggests that the same might be true for the computation of derivatives of the coherent states overlap integral for some coherent states.
The partial derivative approach is commonly used in the computation of the overlap of vibrational
states~\cite{doktorov:1977}, and it is easy to see that the space of $N$ photons in $M$ optical modes is isomorphic
to the space of $N$ molecular vibrational quanta (phonons) in $M$ vibrational modes.
This connection suggests that the dynamics of vibrational modes is computationally
difficult, at least for some instances. However, a simple unitary transformation of the modes is not
sufficient to reproduce vibronic spectra. Additional effects need to be taken into account, as
explained in the next subsection.

\subsection{Vibronic Transitions}
Molecular vibronic spectroscopies such as absorption, emission, photoelectron, and
resonance Raman are fundamental probes for molecular properties.  
The corresponding vibronic transitions involve two electronic states and one can extract the molecular structural and force field changes from the spectra.    
The linear absorption spectra of molecules determines important properties such as their performance as solar-cells~\cite{Johannes2011} or as dyes for either industrial processes~\cite{Gross2000} or biological labels~\cite{Winkler29032013}. 
The prediction of the linear absorption of molecules is challenging computationally, especially when
complicated vibrational features (see \emph{e.g.}~\cite{Dierksen2004,Hayes2011}) make the spectra very rich. 
Photoelectron spectroscopy is a useful tool to study the ionized states of molecules. 
The ionizing process is important in chemistry and biology for example the photodamage  of deoxyribonucleic acid (DNA) molecules is fatal to life. We show a photoelectron spectrum of thymine~\cite{choi:2005} (the experimental spectrum can also be found in Fig.~\ref{fig:thymine} as well) as an example of the current state of the art.  

An electronic transition of a molecule induces nuclear structural and force changes at the new
electronic state. This defines a new set of vibrational modes that are displaced, distorted--hence
showing a frequency change, and rotated with respect to the vibrational modes of the ground
electronic state (see Table~\ref{tab:bsvib} first row and second column). 
Within the harmonic approximation of the electronic energy surfaces and the assumption of 
coordinate-independent electronic transition moment (the Condon approximation), the vibronic transition profiles
can be obtained by the overlap integral of the two $M$-dimensional quantum harmonic oscillator (QHO)
eigenstates (FC integral), where $M=3M_{\mathrm{atom}}-6(5)$ for non-linear (linear) molecules with
$M_{\mathrm{atom}}$ atoms.

In order to describe these effects and compute vibronic profiles, Duschinsky~\cite{duschinsky:1937}
proposed a linear relation between the initial (mass-weighted) normal coordinates ($\mathbf{q}$) and
the final coordinates ($\mathbf{q}'$), which reads

\begin{align}
\mathbf{q}'=\mathbf{U}\mathbf{q}+\mathbf{d} \, ,
\label{eq:duschinsky}
\end{align}
where $\mathbf{U}$ is the Duschinsky rotation (real) matrix and $\mathbf{d}$ is the displacement
(real) vector. $\mathbf{d}$ is responsible for the molecular structural changes along the normal
coordinates. See the first row of Table~\ref{tab:bsvib} for a comparison between the Duschinsky
relation and the boson sampling problem. Observe that all matrices and vectors associated to
the electronic excitation of a molecule are real matrices and real vectors, and this fact will be
used to simplify all the expressions reported below.
The two sets of QHOs are related by the Duschinsky relation and this relation can be expressed in
terms of a modification of the ladder operators~\cite{malmqvist:1998} as given by

\begin{align}
\mathbf{\hat{a}}^{'\dagger}=\frac{1}{2}\left(\mathbf{J}-(\mathbf{J}^{\mathrm{t}})^{-1}\right)\mathbf{\hat{a}}
    +\frac{1}{2}\left(\mathbf{J}+(\mathbf{J}^{\mathrm{t}})^{-1}\right)\mathbf{\hat{a}}^{\dagger}+\frac{1}{\sqrt{2}}\boldsymbol{\delta} \, ,
    \label{eq:duschinskya}
\end{align}

with $\mathbf{J}$ and $\boldsymbol{\delta}$ defined as follow

\begin{align}
&\mathbf{J}=\boldsymbol{\Omega}'\mathbf{U}\boldsymbol{\Omega}^{-1}, \quad
\boldsymbol{\delta}=\hbar^{-\tfrac{1}{2}}\boldsymbol{\Omega}'\mathbf{d}, \nonumber \\
&\boldsymbol{\Omega}'=\mathrm{diag}(\sqrt{\omega_{1}'},\ldots,\sqrt{\omega_{N}'}), \quad 
\boldsymbol{\Omega}=\mathrm{diag}(\sqrt{\omega_{1}},\ldots,\sqrt{\omega_{N}}) \, .
\label{eq:parameters}
\end{align}

The notation ``diag'' denotes a diagonal matrix, while $\omega_{k}'$ and $\omega_{l}$
are the harmonic angular frequencies of  the final and initial states. The major differences
of Eq.~\eqref{eq:duschinskya} from Eq.~\eqref{eq:operatorrotation} are the appearance of the
annihilation operators and the displacement vector $\boldsymbol{\delta}$. The annihilation
operators appear in Eq.~\eqref{eq:duschinskya} to account for the distinct frequencies of
the QHOs. Doktorov \emph{et al.}~\cite{doktorov:1977}  analyzed the linear transformation
in Eq.~\eqref{eq:duschinskya} with a set of unitary operators.  The linear transform in
Eq.~\eqref{eq:duschinskya} can be written as $\mathbf{\hat{a}}^{'\dagger}=\hat{U}_{\mathrm{Dok}}^{\dagger}\mathbf{\hat{a}}^{\dagger}\hat{U}_{\mathrm{Dok}}$,
where the Doktorov transformation $\hat{U}_{\mathrm{Dok}}$ is,

\begin{align}
\hat{U}_{\mathrm{Dok}}&=\hat{D}_{\boldsymbol{\delta}/\sqrt{2}}
\hat{S}_{\boldsymbol{\Omega}'}^{\dagger}\hat{R}_{\mathbf{U}}\hat{S}_{\boldsymbol{\Omega}} \, .
\label{eq:Doktorov}
\end{align}

With our conventions, any initial vibronic state $\ket{\phi_{\text{in}}}$ is transformed
into $\ket{\phi_{\text{out}}}=\hat{U}_{\mathrm{Dok}} \ket{\phi_{\text{in}}}$.
The Doktorov transformation is composed, in order of application, of
(single mode) squeezing $\hat{S}_{\boldsymbol{\Omega}}$,
rotation $\hat{R}_{\mathbf{U}}$,
squeezing $\hat{S}_{\boldsymbol{\Omega}'}^{\dagger}$, and 
coherent state displacement $\hat{D}_{\boldsymbol{\delta}/\sqrt{2}}$ operators.
The specific form of the unitary
operators, together with Eq.~\eqref{eq:operatorrotation}, is given in Ref.~\cite{Ma1990}
and also in the supporting information (SI). 

Accordingly, the transition amplitude in the Duschinsky relation can be obtained from the
partial derivatives of $\mathrm{exp}(\tfrac{1}{2}(\vert\boldsymbol{\gamma}\vert^{2}+\vert\boldsymbol{\alpha}\vert^{2}))\langle\boldsymbol{\gamma};\mathrm{out}\vert\boldsymbol{\alpha};\mathrm{in}\rangle$, \emph{i.e.}

\begin{align}
&\langle\mathbf{m};\mathrm{out}\vert \mathbf{n};\mathrm{in}\rangle=
\langle\mathbf{0};\mathrm{out}\vert\mathbf{0};\mathrm{in}\rangle
\left(\prod_{k=1}^{N}\frac{\partial_{\alpha_{k}}^{n_{k}}\partial_{\gamma_{k}^{*}}^{m_{k}}}{\sqrt{n_{k}! m_{k}!}}\right)
\nonumber \\ 
&\mathrm{e}^{-\tfrac{1}{2}(\boldsymbol{\alpha}^{\mathrm{t}}~\boldsymbol{\gamma}^{\dagger})\mathbf{W}
\left(\begin{smallmatrix}
\boldsymbol{\alpha}\\
\boldsymbol{\gamma}^{*}
\end{smallmatrix}\right)
+\mathbf{r}^{\mathrm{t}}
\left(\begin{smallmatrix}
\boldsymbol{\alpha}\\
\boldsymbol{\gamma}^{*}
\end{smallmatrix}\right)}
\Biggr\rvert_{\boldsymbol{\alpha},\boldsymbol{\gamma}^{*}=\mathbf{0}}  .
\label{eq:MHP}
\end{align}
Here 
$\ket{\boldsymbol{\gamma};\mathrm{out}}=\hat{U}_{\mathrm{Dok}}^{\dagger}\ket{\boldsymbol{\gamma};\mathrm{in}}$ and $\ket{\mathbf{m};\mathrm{out}}=\hat{U}_{\mathrm{Dok}}^{\dagger}\ket{\mathbf{m};\mathrm{in}}$, 
while $\mathbf{W}$ is a $2M\times2M$ matrix and $\mathbf{r}$ is a $2M$-dimensional vector~\cite{doktorov:1977,jankowiak:2007}, which also can be found in SI.

Unlike the usual boson sampling case, the total number of phonons is not conserved in the
scattering process. The calculation of FC integrals in Eq.~\eqref{eq:MHP} is equivalent to the
evaluation of multivariate Hermite polynomials at the origin~\cite{Huh2011a,Huh2011,Huh2012}.
Indeed, Huh~\cite{Huh2011a} showed that Eq.~\eqref{eq:MHP} can be evaluated by an algorithm
developed for multivariate normal moments~\cite{kan:2007}.
Kan~\cite{kan:2007} exploited collective variable to calculate the moments of the distribution,
obtaining an algorithm that requires $(1+[\tfrac{1}{2}(\sum_{k}(n_{k}+m_{k}))])\prod_{k}(n_{k}+1)(m_{k}+1)$ terms, where [x] is a rounded integer of x. 
This number is much smaller than the number of terms from a brute force
evaluation of the Wick's formula, corresponding to ($\sum_{k}(n_{k}+m_{k})-1)!!$~\cite{kan:2007}, 
where a similar analysis was done for the squeezed vacuum state input problem in boson
sampling~\cite{rahimi2014}. However, the computation with Kan's algorithm~\cite{kan:2007} still likely to 
be a hard problem.     

The transition probability ($\vert\langle\mathbf{m};\mathrm{out}\vert \mathbf{n};\mathrm{in}\rangle\vert^{2}$)
is called the Franck-Condon factor and the Franck-Condon profile (FCP) is given at the
vibrational transition frequency ($\omega_{\mathrm{vib}}$). The FCP at $0$K is obtained with the initial vacuum state
$\vert\mathbf{0};\mathrm{in}\rangle$ as given by 

\begin{align}
\mathrm{FCP}(\omega_{\mathrm{vib}})=\sum_{\mathbf{m}}^{\boldsymbol{\infty}}\vert\langle\mathbf{m};\mathrm{out}\vert \mathbf{0};\mathrm{in}\rangle\vert^{2}\delta(\omega_{\mathrm{vib}}-\sum_{k}^{N}\omega_{k}'m_{k}) \, .
\label{eq:FCP}
\end{align}

Although no rigorous proofs of the complexity of computing FCPs exist, we describe
the observed computational effort for current algorithms and typical (molecular)
problem instances. As the molecular system size and temperature increase, the
evaluation of the FCP with classical computers becomes practically intractable (\emph{cf.} Refs.~\cite{jankowiak:2007,santoro:2007b}).  
The size and temperature effects make the resulting spectrum very congested due to
the increase of the density of states. Already the enumeration of the states
contributing to each point of the frequency grid ($\omega_{\mathrm{vib}}$) is an issue to evaluate the FCP. 
That is, one needs to find all sets of $\mathbf{m}$ that satisfy $\omega_{\mathrm{vib}}=\sum_{k}^{N}\omega_{k}'m_{k}$ at $0$K.
To address this issue, one should find an algorithm to count the vibrational states
and determines its limitation with respect to the system size, see for example
Ref.~\cite{berger:1997}.

We summarize the comparison between the boson sampling and the vibronic transition
in Table~\ref{tab:bsvib} and proceed to show how to simulate the molecular vibronic
spectra by sampling photons from a modified boson sampling device.

\section{Boson Sampling for Franck-Condon Factors}
If all the phonon frequencies are identical and there is no displacement, the Duschinsky
relation (Eq.~\eqref{eq:duschinskya}) can be directly reduced to the original boson sampling
problem (Eq.~\eqref{eq:operatorrotation}). Therefore the Duschinsky relation can be considered
as a generalized boson sampling problem (\emph{cf.} Ref.~\cite{PhysRevLett.113.100502})
which involves not only rotation but also displacement and squeezing operations.   
In this section, we modify boson sampling to simulate the FCP in the Duschinksy relation. 
We assume that the initial state corresponds to the vibrational ground state (mathematically,
a vacuum state), which means that the FCP is produced at 0K. 
Our proposal can be extended to vibronic profiles at finite temperature by preparing various
initial states with a probability that corresponds to their Boltzmann
factor~\cite{santoro:2008,Huh2011a}. A detailed finite-temperature experimental proposal is outside the scope of this paper.  
We can interpret some of the additional operators in the Duschinsky relation as part of
the state preparation process of the input state for boson sampling. To this end, we move
the position of the displacement operator in $\hat{U}_{\mathrm{Dok}}$ (Eq.~\eqref{eq:Doktorov})
from the left end to the right end by rotating the corresponding displacement parameter vector,
\emph{i.e.}

\begin{align}
\hat{U}_{\mathrm{Dok}}=
\hat{S}_{\boldsymbol{\Omega}'}^{\dagger}\hat{R}_{\mathbf{U}}\hat{S}_{\boldsymbol{\Omega}}\hat{D}_{\mathbf{J}^{-1}\boldsymbol{\delta}/\sqrt{2}} \, .
\label{eq:Doktorov2}
\end{align}

\begin{figure}[htb]
\begin{center}
\includegraphics[width=\linewidth]{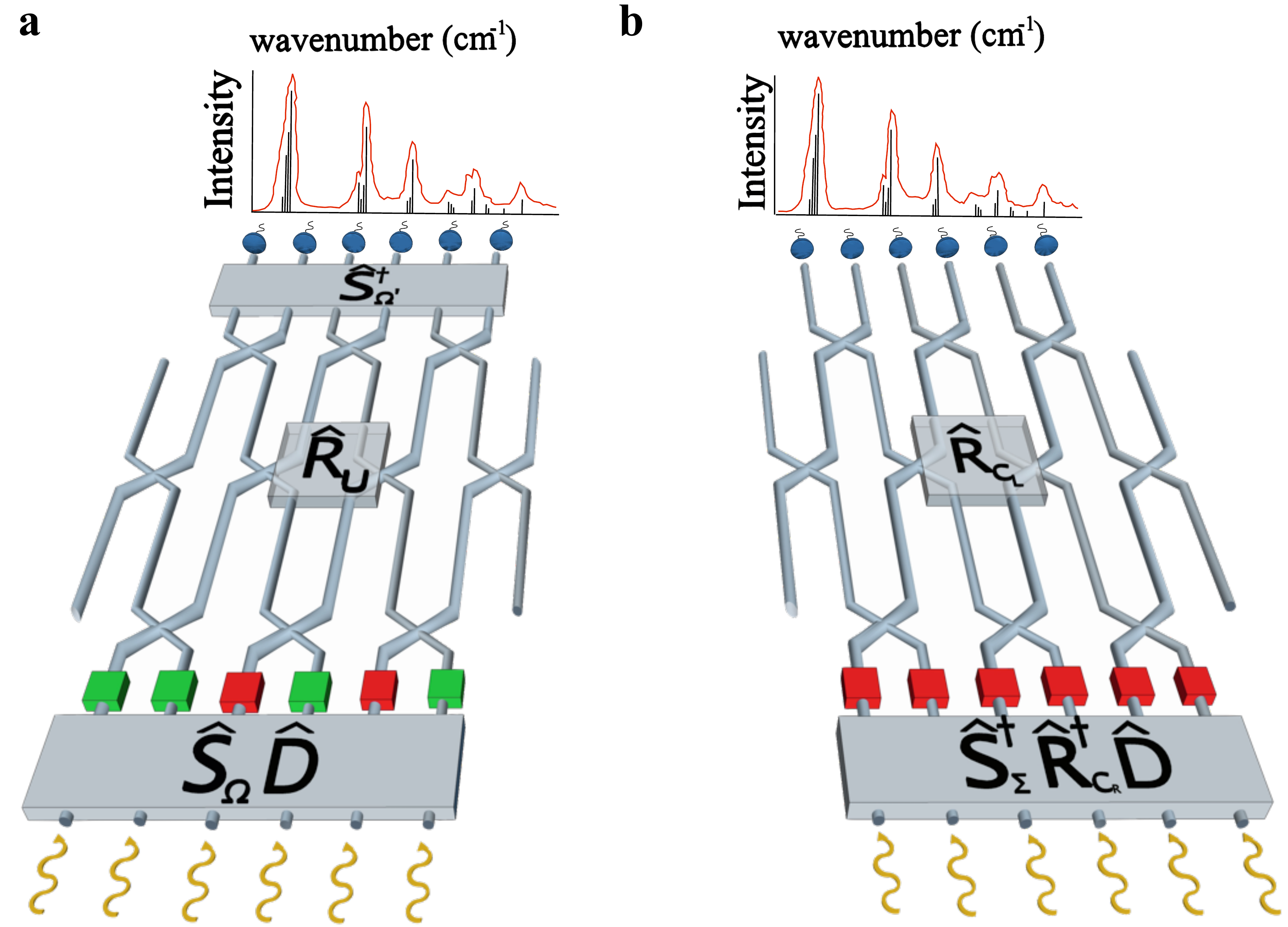}
\caption{Boson Sampling apparatus for Vibronic Spectra. {\bf a}, The boson sampling
apparatus modified according to a direct implementation of Eq.~\eqref{eq:Doktorov2}.
{\bf b}, The boson sampling apparatus modified according to Eq.~\eqref{eq:Doktorov3}.
Here the difference with the usual setups for the typical boson sampling problem
is confined to the preparation process of the input state. For simplification,
$\hat{D}=\hat{D}_{\mathbf{J}^{-1}\boldsymbol{\delta}/\sqrt{2}}$. 
Green and red boxes after the first unitary operations represent the prepared initial states which are identified as squeezed vacuum and squeezed coherent states respectively.
}\label{fig:bsvibapp}
\end{center}
\end{figure}

The Franck-Condon optical apparatus can be set up according to $\hat{U}_{\mathrm{Dok}}$
in Eq.~\eqref{eq:Doktorov2}.  As shown in Fig.~\ref{fig:bsvibapp}{\bf a}, the photons are
prepared as squeezed coherent states or squeezed vacuum states, which correspond to the
displaced modes and non-displaced modes respectively. Thus, the input state to the boson
sampling optical network is
$\vert\psi\rangle=\hat{S}_{\boldsymbol{\Omega}}\hat{D}_{\mathbf{J}^{-1}\boldsymbol{\delta}/\sqrt{2}}\vert \mathbf{0};\mathrm{in}\rangle=\hat{S}_{\boldsymbol{\Omega}}\vert \tfrac{1}{\sqrt{2}}\mathbf{\mathbf{J}^{-1}\boldsymbol{\delta}};\mathrm{in}\rangle$.
As depicted in Fig.~\ref{fig:bsvibapp}{\bf a}, the prepared initial state $\vert\psi\rangle$
passes through the boson sampling photon scatterer $\hat{R}_{\mathbf{U}}$ and then the output
photons undergo the second squeezing operation $\hat{S}_{\boldsymbol{\Omega}'}^{\dagger}$. Finally,
photocounters detect the output Fock states. The resulting probability can be resolved in
its transition frequency ($\omega_{\mathrm{vib}}=\sum_{k}^{N}\omega_{k}'m_{k}$) to yield
the FCPs from the boson sampling statistics. We note that, here, $\omega_{k}'$ represents
the phonon frequency and not the input photon frequency. We do not assign different
frequencies to different modes for the corresponding phonon modes, however the fact that
phonon modes frequencies are different is taken into account by parameters of the state
preparation process and of the optical network. 

In practice, the second squeezing operation is difficult to realize in optical setups since
one needs a non-linear interaction in situations that may involve only a limited number of
photons. For this reason, instead of performing such operation directly, as described in
Fig.~\ref{fig:bsvibapp}{\bf a}, we propose to compress the two squeezing operations into
a single one. We can achieve this goal by means of the singular value decomposition (SVD)
of the matrix $\mathbf{J}$ in Eq.~\eqref{eq:parameters}, 

\begin{align}
\mathbf{J}=\mathbf{C}_{\mathrm{L}}\boldsymbol{\Sigma}\mathbf{C}_{\mathrm{R}}^{\mathrm{t}}\, ,
\end{align}
where $\mathbf{C}_{\mathrm{L}}$ and $\mathbf{C}_{\mathrm{R}}$ are real unitary matrices and
$\boldsymbol{\Sigma}$ is a diagonal matrix composed of square roots of the eigenvalues of
$\mathbf{J}^{\mathrm{t}}\mathbf{J}$. As a result the Doktorov operator can be rewritten as

\begin{align}
\hat{U}_{\mathrm{Dok}}=
\hat{R}_{\mathbf{C}_{\mathrm{L}}}\hat{S}_{\boldsymbol{\Sigma}}^{\dagger}\hat{R}_{\mathbf{C}_{\mathrm{R}}}^{\dagger}\hat{D}_{\tfrac{1}{\sqrt{2}}\mathbf{J}^{-1}\boldsymbol{\delta}} \, .
\label{eq:Doktorov3}
\end{align}

At this point, the Doktorov operator is composed of two rotations, one squeezing operator,
and one displacement operator. The input state $\vert\phi\rangle$ to the boson sampling
optical network is prepared by applying the displacement, rotation and squeezing operators
sequentially, \emph{i.e.}

\begin{align}
\vert\phi\rangle&=\hat{S}_{\boldsymbol{\Sigma}}^{\dagger}\hat{R}_{\mathbf{C}_{\mathrm{R}}}^{\dagger}\hat{D}_{\mathbf{J}^{-1}\boldsymbol{\delta}/\sqrt{2}} \vert\mathbf{0};\mathrm{in}\rangle \\ \nonumber
&=\hat{S}_{\boldsymbol{\Sigma}}^{\dagger}\vert\tfrac{1}{\sqrt{2}}\mathbf{\mathbf{C}_{\mathrm{R}}^{\mathrm{t}}\mathbf{J}^{-1}\boldsymbol{\delta}};\mathrm{in}\rangle \, .
\end{align}

As one can see from direct inspection, $\vert\phi\rangle$ is a squeezed coherent state.
The only remaining task is to pass the prepared input state through the boson sampling
optical network, which is characterized by the rotation matrix $\mathbf{C}_{\mathrm{L}}$
for $\hat{R}_{\mathbf{C}_{\mathrm{L}}}$. 
This simplified optical apparatus is depicted in Fig.~\ref{fig:bsvibapp}{\bf b}.
Now, the problem is identical to the boson sampling with squeezed coherent states
as input~\cite{PhysRevLett.113.100502,rahimi2014,olson2014}. Boson sampling with inputs different from
Fock states, 
for example with coherent states or squeezed vacuum states, have been proposed
and analysed in the context of the study of computational complexity in
Refs.~\cite{PhysRevLett.113.100502,rahimi2014,olson2014}.

\section{Examples}

\begin{figure}[htb]
\begin{center}
\includegraphics[width=\linewidth]{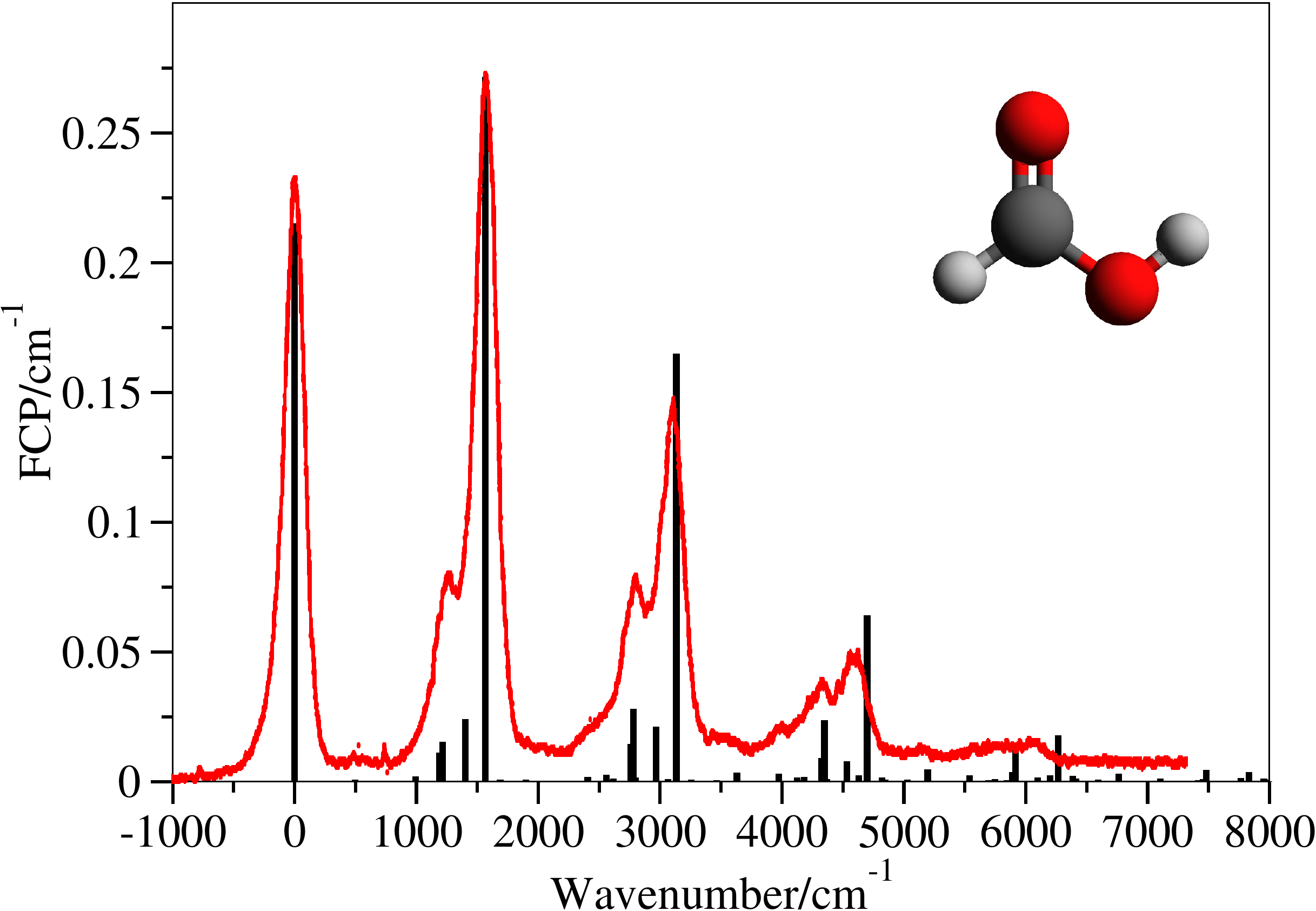}
\caption{ Franck-Condon profile (black sticks) of formic acid ($1~^{1}\mathrm{A}^{'}\rightarrow1~^{2}\mathrm{A}^{'}$) for a symmetry block $\mathrm{a}^{'}$. The red curve is taken from the experimental spectrum in Leach \emph{et al.}~\cite{leach:2003}. 
}\label{fig:formicacid}
\end{center}
\end{figure}

\begin{figure}[htb]
\begin{center}
\includegraphics[width=\linewidth]{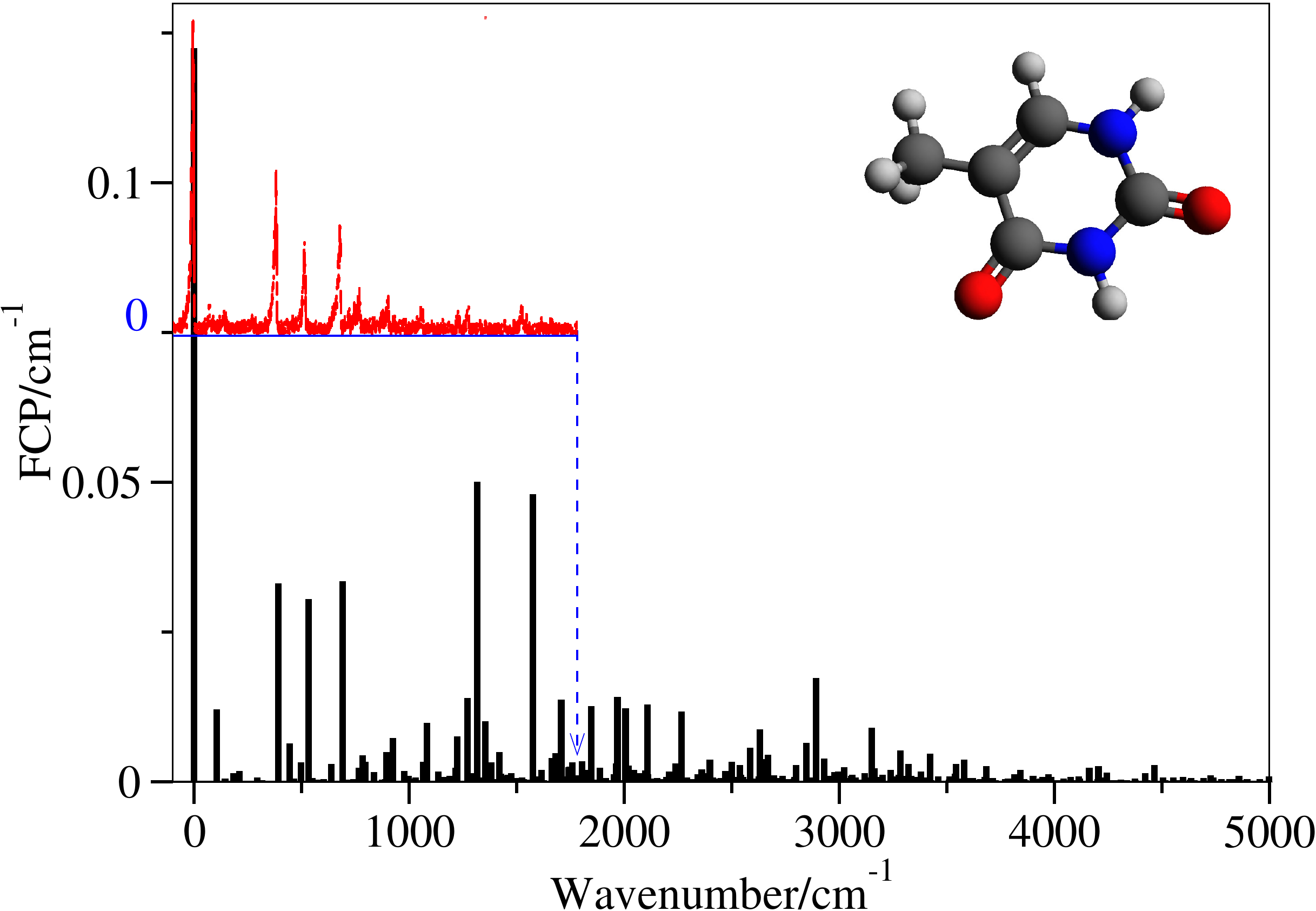}
\caption{Franck-Condon profile (black sticks) of thymine  ($1~^{1}\mathrm{A}^{'}\rightarrow1~^{2}\mathrm{A}^{''}$). 
The red curve, whose Franck-Condon profile is shifted to be compared clearly, is taken from the experimental spectrum in Choi \emph{et al.}~\cite{choi:2005}.
}\label{fig:thymine}
\end{center}
\end{figure}

We present two examples of computation of the FC factors for molecules. In particular,
we propose to simulate the photoelectron spectra of formic acid (CH$_{2}$O$_{2}$) and
thymine (C$_{5}$H$_{6}$N$_{2}$O$_{2}$). 
The photoelectron spectroscopy involves the molecular electronic transition from a
neutral state to a cationic state. The spectral profile can be obtained by computing
the corresponding FC factors~\cite{jankowiak:2007}. 
The molecular parameters for the calculations are reproduced from the supplementary
material of Jankowiak \emph{et al.}~\cite{jankowiak:2007}. The FCPs are calculated
with the vibronic structure program {\bf hotFCHT}~\cite{berger:1997,berger:1998,jankowiak:2007}. 
Parameters for the corresponding boson sampling experimental setup are given in SI.

Formic acid represents a small system to test the quantum simulation with relatively
small optical setups. 
The calculated FCP for the formic acid are presented in Fig.~\ref{fig:formicacid} as black sticks,
with a bin size $\Delta_\mathrm{vib}=1$ cm$^{-1}$. 
The red curve in Fig.~\ref{fig:formicacid} is taken from the experimental spectrum in Leach
\emph{et al.}~\cite{leach:2003} and includes effects of line broadening. A table for 
the probabilities with respect to the corresponding quantum numbers and the vibrational
transition frequencies are given in SI for direct verification with
boson sampling experiments. Additionally, we simulate the results of what one would expect in a boson sampling simulation of Formic acid, and present in SI. This is done by stochastically sampling the known probability distribution for the output modes and perform the analysis according to Eq.~\eqref{eq:FCP}. The results from this simulation indicate relatively few samples are needed from the device to resolve the overall shape of the spectrum, supporting the experimental feasibility of the approach.

The important FC factors ($\ge 0.01$) in Fig.~\ref{fig:formicacid} require at most 3 photons 
per mode (see SI). Current photon counters are able to distinguish up to a few photon numbers ($\le 3$) per mode~\cite{Carolan2014}. 
The (single mode) squeezing parameters for formic acid are given as $\mathrm{ln}(\mathbf{\Sigma})$~\cite{Ma1990}, that is,  
\begin{align}
\mathrm{ln}(\boldsymbol{\Sigma})=
\mathrm{diag}\left(
0.10,    0.07,    0.02,   -0.06,   -0.08,   -0.11,   -0.19
\right) .
\end{align}
The squeezing parameters are between -0.2 and 0.1. 
These parameters are related to the frequency ratio between 
the initial and final frequencies. 
The experimental implementation of boson sampling for vibronic spectra would rely on experimental 
squeezing operation techniques. At present, multiple experimental proposals for arbitrary squeezing operations on coherent states have been proposed. For example, phase intensive optical amplification~\cite{Josse2006}, ancillary squeezed vacuum~\cite{Yoshikawa2007}, and dynamic squeezing operation~\cite{Miwa2014} are all potentially promising. 

We present here an FCP of thymine as a more experimentally challenging example for vibronic boson sampling.  
The calculated FCP of thymine is given as black sticks in Fig.~\ref{fig:thymine}. The details can be found in Ref.~\cite{jankowiak:2007} and also in SI.

\section{Outlook and conclusions}

Boson sampling may be one of the first experimentally accessible systems that challenges the
computational power of classical computers.  However, to our knowledge there has not been any proposal on its use for simulation purposes. 
In this work, we develop a connection between 
molecular vibronic spectra and boson sampling that allows the calculation of Franck-Condon
profiles with quantum optical networks. First, we note that the such connection suggests
that computing the dynamics of vibrational systems must be a computationally difficult
task for some systems. Then, we show that a modification  of the input
state of a boson sampling device enables the computation of complex molecular spectra
in a way that includes effects beyond simple vibrational dynamics.
This allows one to generate the molecular vibronic spectrum by shining light in boson
sampling optical networks rather than on real molecules, and opens new possibilities for
studying molecules that are hard to isolate in a lab setting and too big to simulate on
a classical computer.  

To motivate experimental realizations, we present two small molecules having C$_{\mathrm{s}}$
point group symmetry. Exploiting the molecular symmetry makes the classical computation of
the FCP easier, but molecules have often no symmetry, especially in case of large molecules.
Testing small systems represents an important step preliminary to the application of boson
sampling to molecular vibronic spectroscopy of large systems whose calculation with
classical computers is expected to be hard.
Our work can be extend in various directions: For example, the quantum simulation that we
propose can be generalized to vibronic profile at finite temperature~\cite{santoro:2008,Huh2011a} by exploiting
thermal coherent states~\cite{rahimi2014} or one can consider the modification of boson
sampling experiments to include non-Condon~\cite{santoro:2008} and anharmonic effects~\cite{huh:2010anharm}.

\section*{Acknowledgments}
We thank Prof. Dr. Robert Berger for permission to use the vibronic structure program {\bf hotFCHT} for our research. J. H. and A. A.-G. acknowledge Defense Threat Reduction Agency grant HDTRA1-10-1-0046 and the Air Force Office of Scientific Research grant FA9550-12-1-0046. J. M. is supported by the Department of Energy Computational Science Graduate Fellowship under grant number DE-FG02-97ER25308. G. G. G. and A. A.-G. acknowledge support from NSF Grant No. CHE-1152291. B. P. and A. A.-G acknowledge support from the STC Center for Integrated Quantum Materials, NSF Grant No. DMR-1231319. Further, A. A.-G. is grateful for the support from Defense Advanced Research Projects Agency grant N66001-10-1-4063, and the Corning Foundation for their generous support.

\section{Author contributions}
J. H., G. G. G. and A. A.-G. designed the research; all authors were involved in the
research process, in discussions and in writing the paper.

\section{Competing financial interests}
The authors declare no competing financial interests.


\appendix
\renewcommand\thefigure{\thesection.\arabic{figure}}
\renewcommand\thetable{\thesection.\arabic{table}}
\setcounter{figure}{0} 
\setcounter{table}{0} 

\section{Supporting information}
\subsection{Duschinsky}
The specific form of the unitary
operators, together with Eq.~\eqref{eq:operatorrotation}, is given by \cite{Ma1990}, 

\begin{align}
&\hat{D}_{\boldsymbol{\delta}/\sqrt{2}}^{\dagger}\mathbf{\hat{a}}^{\dagger}\hat{D}_{\boldsymbol{\delta}/\sqrt{2}}=\mathbf{\hat{a}}^{\dagger}+\tfrac{1}{\sqrt{2}}\boldsymbol{\delta}, \\
&\hat{S}_{\boldsymbol{\Omega}'}\mathbf{\hat{a}}^{\dagger}\hat{S}_{\boldsymbol{\Omega}'}^{\dagger}=\tfrac{1}{2}\left(\boldsymbol{\Omega'}+\boldsymbol{\Omega'}^{-1}\right)\mathbf{\hat{a}}^{\dagger}
+\tfrac{1}{2}\left(\boldsymbol{\Omega'}-\boldsymbol{\Omega'}^{-1}\right)\mathbf{\hat{a}}, \\
&\hat{S}_{\boldsymbol{\Omega}}^{\dagger}\mathbf{\hat{a}}^{\dagger}\hat{S}_{\boldsymbol{\Omega}}=\tfrac{1}{2}\left(\boldsymbol{\Omega}+\boldsymbol{\Omega}^{-1}\right)\mathbf{\hat{a}}^{\dagger}
-\tfrac{1}{2}\left(\boldsymbol{\Omega}-\boldsymbol{\Omega}^{-1}\right)\mathbf{\hat{a}}\, .
\end{align}

The closed form of the integral of the two coherent states in the
Duschinsky relation can be obtained as~\cite{doktorov:1977,jankowiak:2007}

\begin{align}
&\langle\boldsymbol{\gamma};\mathrm{out}\vert\boldsymbol{\alpha};\mathrm{in}\rangle=\langle\boldsymbol{\gamma};\mathrm{in}\vert \hat{U}_{\mathrm{Dok}}\vert\boldsymbol{\alpha};\mathrm{in}\rangle\\ \nonumber
&=\langle\mathbf{0};\mathrm{out}\vert\mathbf{0};\mathrm{in}\rangle\mathrm{e}^{-\tfrac{1}{2}(\vert\boldsymbol{\gamma}\vert^{2}+\vert\boldsymbol{\alpha}\vert^{2})}
\mathrm{e}^{-\tfrac{1}{2}(\boldsymbol{\alpha}^{\mathrm{t}}~\boldsymbol{\gamma}^{\dagger})\mathbf{W}
\left(\begin{smallmatrix}
\boldsymbol{\alpha}\\
\boldsymbol{\gamma}^{*}
\end{smallmatrix}\right)
+\mathbf{r}^{\mathrm{t}}
\left(\begin{smallmatrix}
\boldsymbol{\alpha}\\
\boldsymbol{\gamma}^{*}
\end{smallmatrix}\right)} \, .
\label{eq:FCI}
\end{align}

$\mathbf{W}$ is a self-inverse $2M\times 2M$ matrix and $\mathbf{r}$ is a
$2M$-dimensional vector, which are defined as follows,

\begin{align}
\mathbf{W}=
\begin{pmatrix}
\mathbf{I}-2\mathbf{Q} & -2\mathbf{R}\\
-2\mathbf{R}^{\mathrm{t}} & \mathbf{I}-2\mathbf{P}
\end{pmatrix} , \quad
\mathbf{r}=\sqrt{2}
\begin{pmatrix}
-\mathbf{R}\boldsymbol{\delta}\\
(\mathbf{I}-\mathbf{P})\boldsymbol{\delta}
\end{pmatrix}, 
\end{align}

where $\mathbf{I}$ is a $M$-dimensional identity matrix,
$\mathbf{Q}=(\mathbf{I}+\mathbf{J}^{\mathrm{t}}\mathbf{J})^{-1}$,
$\mathbf{P}=\mathbf{J}\mathbf{Q}\mathbf{J}^{\mathrm{t}}$ and $\mathbf{R}=\mathbf{Q}\mathbf{J}^{\mathrm{t}}$. 
The overlap integral of the two vacuum states in the Duschinsky relation (Eq.~\eqref{eq:duschinsky}
and~\ref{eq:duschinskya}) are given as, 

\begin{align}
\langle\mathbf{0};\mathrm{out}\vert\mathbf{0};\mathrm{in}\rangle=2^{\tfrac{N}{2}}\vert\det(\mathbf{R})\vert^{\tfrac{1}{2}}\mathrm{e}^{-\tfrac{1}{2}
\boldsymbol{\delta}^{\mathrm{t}}(\mathbf{I}-\mathbf{P})\boldsymbol{\delta}}.
\end{align}

$\mathbf{J}$ and $\boldsymbol{\delta}$ are defined as follow

\begin{align}
&\mathbf{J}=\boldsymbol{\Omega}'\mathbf{U}\boldsymbol{\Omega}^{-1}, \quad
\boldsymbol{\delta}=\hbar^{-\tfrac{1}{2}}\boldsymbol{\Omega}'\mathbf{d}, \nonumber \\
&\boldsymbol{\Omega}'=\mathrm{diag}(\sqrt{\omega_{1}'},\ldots,\sqrt{\omega_{N}'}), \quad 
\boldsymbol{\Omega}=\mathrm{diag}(\sqrt{\omega_{1}},\ldots,\sqrt{\omega_{N}}) \, .
\end{align}

\subsection{Formic acid}
Franck-Condon factors ($\ge 0.01$) are listed in Table~\ref{tab:formicacid}.
\begin{table}[htb]
  \centering
  \ra{1.3}
\begin{tabular}{@{}cc@{}}
\hline\\
Frequency ($\omega_{\mathrm{vib}}$)/cm$^{-1}$ & Franck-Condon factor\\
\hline\\
0 & 0.2152\\
$\omega_{3}'$ & 0.2717\\
$2\omega_{3}'$ & 0.1649\\
$3\omega_{3}'$ & 0.0640\\
$4\omega_{3}'$ & 0.0178\\
$\omega_{3}'+\omega_{4}'$ & 0.0211\\
$\omega_{3}'+\omega_{5}'$ & 0.0281\\
$\omega_{3}'+\omega_{6}'$ & 0.0145\\
$2\omega_{3}'+\omega_{5}'$ & 0.0237\\
$3\omega_{3}'+\omega_{5}'$ & 0.0123\\
$\omega_{4}'$ & 0.0242\\
$\omega_{5}'$ & 0.0153\\
$\omega_{6}'$ & 0.0112\\
\hline
\end{tabular}
\caption{Franck-Condon factors ($\ge 0.01$) in Fig.~\ref{fig:formicacid}}
\label{tab:formicacid}
\end{table} 
The photoelectron spectrum of formic acid ($1~^{1}\mathrm{A}^{'}\rightarrow1~^{2}\mathrm{A}^{'}$)
can be found in Leach \emph{et al.}~\cite{leach:2003}.
Formic acid has a C$_{\mathrm{s}}$ point group symmetry, which is composed of two
irreducible representations $\mathrm{a}^{'}$ and $\mathrm{a}^{''}$. 
As a result the Duschinsky rotation matrix for formic acid is in a block diagonal form
with the blocks corresponding to the  irreducible representations. 
The overall spectral shape is dominated by the contribution from the totally symmetric
block $\mathrm{a}^{'}$ while the non-totally symmetric block $\mathrm{a}^{''}$ only provides
a minor correction to the spectral shape due to its lack of the structural displacements.
Therefore, here we give the characteristic matrices and vectors for the symmetry block
$\mathrm{a}^{'}$, which is composed of 7 QHOs for a simple boson sampling experiment.

Below, we provide the characteristic parameters for formic acid for its symmetric block $a'$ as

\begin{align}
&\mathbf{U}=
\left(
\scalefont{0.9}{
\begin{smallmatrix}
 0.9934  &  0.0144 &   0.0153  &  0.0286 &   0.0638 &   0.0751 &  -0.0428\\
   -0.0149  &  0.9931 &   0.0742  &  0.0769 &  -0.0361 &  -0.0025 &   0.0173\\
   -0.0119  & -0.0916 &   0.8423  &  0.1799 &  -0.3857 &   0.3074 &   0.0801\\
    0.0381  &  0.0409 &  -0.3403  & -0.5231 &  -0.6679 &   0.3848 &   0.1142\\
   -0.0413  & -0.0342 &  -0.4004  &  0.7636 &  -0.1036 &   0.4838 &   0.0941\\
    0.0908  & -0.0418 &  -0.0907  &  0.3151 &  -0.5900 &  -0.7193 &   0.1304\\
   -0.0325  &  0.0050 &  -0.0206  &  0.0694 &  -0.2018 &   0.0173 &  -0.9759
\end{smallmatrix}
}
\right),\\
&\boldsymbol{\omega}=
\left(
\scalefont{0.9}{
\begin{smallmatrix}
3765.2386\\ 3088.1826\\ 1825.1799\\ 1416.9512\\ 1326.4684\\ 1137.0490\\ 629.7144
\end{smallmatrix}
}
\right),\quad 
\boldsymbol{\omega}'=
\left(
\scalefont{0.9}{
\begin{smallmatrix}
3629.9472 \\3064.9143\\ 1566.4602\\ 1399.6554\\ 1215.3421\\ 1190.9077\\ 496.2845
\end{smallmatrix}
}
\right),\quad
\boldsymbol{\delta}=
\left(
\scalefont{0.9}{
\begin{smallmatrix}
0.2254   \\ 0.1469 \\   1.5599 \\  -0.3784 \\   0.4553 \\  -0.3439 \\   0.0618
\end{smallmatrix}
}
\right)
\end{align}

where the vibrational frequencies are in cm$^{-1}$, while the other quantities are dimensionless. 

The corresponding boson sampling parameters, can be obtained
by singular value decomposition of the $\mathbf{J}$ matrix, \emph{i.e.} $\mathbf{J}=\mathbf{C}_{\mathrm{L}}\boldsymbol{\Sigma}\mathbf{C}_{\mathrm{R}}^{\mathrm{t}}$

\begin{align}
&\mathbf{C}_{\mathrm{L}}=
\left(
\scalefont{0.9}{
\begin{smallmatrix}
   -0.0786 &   0.6624 &  -0.1910 &   0.0194 &  -0.7022  &  0.1170  &  0.1069\\
    0.1918 &  -0.1188 &  -0.8128 &  -0.5265 &   0.0841  &  0.0637  &  0.0039\\
    0.6084 &   0.0851 &  -0.1492 &   0.3436 &  -0.0404  & -0.6888  &  0.0792\\
    0.6373 &  -0.0386 &   0.4649 &  -0.4920 &  -0.2417  &  0.2050  & -0.1839\\
    0.3455 &   0.2308 &  -0.1980 &   0.4883 &   0.2781  &  0.5577  & -0.4017\\
   -0.0348 &  -0.6595 &  -0.1588 &   0.2914 &  -0.6007  &  0.0687  & -0.2968\\
   -0.2454 &   0.2240 &   0.0069 &  -0.1973 &   0.0396  & -0.3874  & -0.8361
\end{smallmatrix}
}
\right),\\
&\boldsymbol{\Sigma}=
\left(
\scalefont{0.9}{
\begin{smallmatrix}
    1.1020  &       0  &       0  &       0  &       0  &       0 &        0\\
         0  &  1.0728  &       0  &       0  &       0  &       0 &        0\\
         0  &       0  &  1.0214  &       0  &       0  &       0 &        0\\
         0  &       0  &       0  &  0.9420  &       0  &       0 &        0\\
         0  &       0  &       0  &       0  &  0.9276  &       0 &        0\\
         0  &       0  &       0  &       0  &       0  &  0.8941 &        0\\
         0  &       0  &       0  &       0  &       0  &       0 &   0.8296
\end{smallmatrix}
}
\right),\\
&\mathbf{C}_{\mathrm{R}}=
\left(
\scalefont{0.9}{
\begin{smallmatrix}
   -0.0691 &   0.5634 &  -0.1635 &   0.0188 &  -0.7859 &   0.1322 &   0.1248\\
    0.1446 &  -0.0943 &  -0.7600 &  -0.6104 &   0.0841 &   0.1129 &   0.0120\\
    0.1759 &   0.0478 &  -0.2556 &   0.1972 &  -0.0150 &  -0.8645 &   0.3390\\
    0.0237 &   0.0326 &  -0.5446 &   0.7256 &   0.1295 &   0.1946 &  -0.3474\\
   -0.6311 &   0.3592 &  -0.1218 &   0.0398 &   0.4392 &   0.1241 &   0.4979\\
    0.6132 &   0.6591 &   0.1373 &  -0.0571 &   0.3982 &   0.0877 &  -0.0333\\
    0.4104 &  -0.3268 &  -0.0111 &   0.2383 &  -0.0825 &   0.4017 &   0.7069
\end{smallmatrix}
}
\right) .
\end{align}

\subsection{Thymine}
The computed vibronic spectrum
in Ref.~\cite{jankowiak:2007} is in a good agreement with the experimental one. Thymine
also belongs to the C$_{\mathrm{s}}$ point group. As a consequence, The Duschinsky rotation
matrix is also in a block diagonal form. There are 26 and 13 vibrational degrees of freedom
for the irreducible representations $\mathrm{a}^{'}$ and $\mathrm{a}^{''}$, respectively.
In Fig.~\ref{fig:thymine} we show the reproduced FCP based on information from
Ref.~\cite{jankowiak:2007} with {\bf hotFCHT}~\cite{berger:1997,berger:1998,jankowiak:2007}.
The characteristic matrices and vectors of each irreducible representation can be provided on the request.
The full spectrum in Fig.~\ref{fig:thymine} can be obtained by convoluting the two FCPs from
each irreducible representation. For the boson sampling experiments, the full spectrum can
be obtained either by performing experiments on the full system in a block diagonal form or
by convoluting the results from each blocks. 

\subsection{Algorithm and scaling}
\begin{figure}[htb]
\begin{center}
\includegraphics[width=\linewidth]{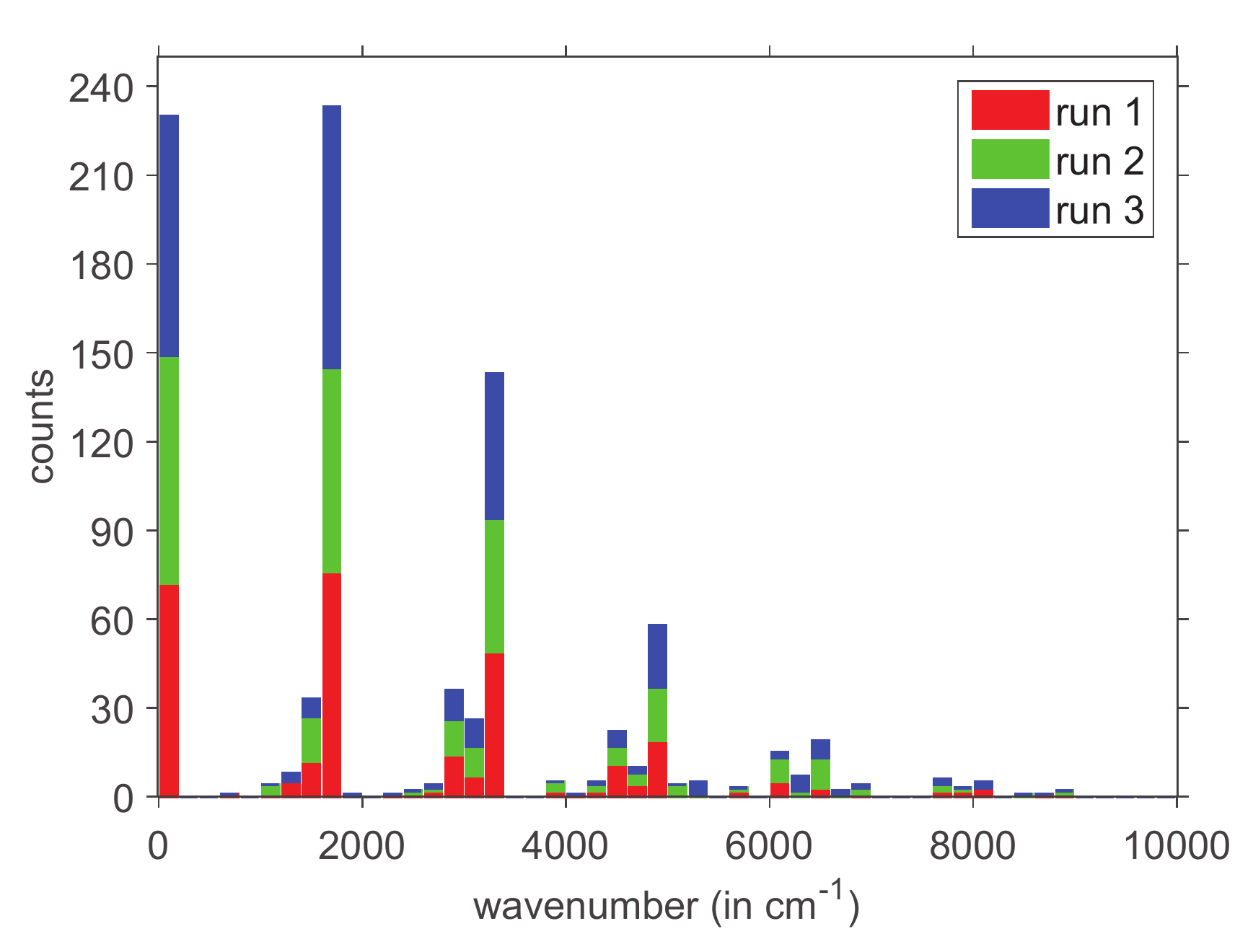}
\caption{ We report the results from a stochastic simulation of the Franck-Condon profile of formic acid according to Eq.~\eqref{eq:FCPsample} (without normalization). The colors correspond to subsequent runs of 300 independent samples each of the modified boson sampling experiment.  This plot demonstrates how few samples one may require from the boson sampling device to resolve a spectra. The bin size in this simulation is set to $\Delta_\mathrm{vib}=200$ cm$^{-1}$.
}\label{fig:formicacid_MC}
\end{center}
\end{figure}

In this section, we explicitly describe the algorithm for computing FCPs given a boson sampling setup
and analyse the computational cost of resolving FCPs with this setup.  As in the rest of the work we limit ourselves to
the case of 0K with the generalization to finite temperature being the subject of future work.

The goal is this case is to resolve the function $\mathrm{FCP}(\omega_{\mathrm{vib}})$ in Eq.~\eqref{eq:FCP} to a fixed precision
$\epsilon_{\mathrm{FCP}}$ in the function value at a fixed resolution $\Delta_\mathrm{vib}$ in the value of 
$\omega_{\mathrm{vib}}$.  We also take as input values the number of vibrational modes $M$, final vibrational frequencies $\{\omega'_k\}$, and a maximum frequency of interest $\omega_{\mathrm{max}}$.

Consider the FCP on the interval $[0, \omega_{\mathrm{max}}]$ and discretize this interval uniformly at a resolution of $\Delta_\mathrm{vib}$.  The algorithm proceeds as follows.  Prepare the state $\ket{\phi}$, and pass it into the boson sampling setup.  Measure at the output modes the photon numbers $\{m_k\}$ in each mode.  Locate the discrete bin of the FCP that is non-zero corresponding to the
the measurement values of $\{m_k\}$, and increment its value by 1.  Repeat the experiment until the estimated statistical error of the average values in each discrete bin of FCP is below the desired threshold $\epsilon_{\mathrm{FCP}}$.  Denote the total number of samples taken as $N_{\mathrm{Samp}}$.

To assess the algorithm, we will rewrite it as stochastic sampling problem over a probability distribution given by the boson sampling device.  We observe that $P_{\mathbf{m \ket{\phi}}} = |\braket{\mathbf{m};\mathrm{out}}{\phi}|^2$ is a normalized probability distribution, and the one naturally sampled at unit cost by a boson sampling device.  As such, an FCP at a given frequency is equivalent to the average value of $f(\mathbf{m}) = \delta(\omega_{\mathrm{vib}}-\sum_{k}^{M}\omega_{k}'m_{k})$ 
over the probability distribution $P$, which we denote $\langle f \rangle_{P}$.  By simply computing the average of $N_{\mathrm{Samp}}$ independent samples $\ket{\mathbf{s}_i}$ taken from the device, that is
\begin{align}
\mathrm{FCP} = \langle f \rangle_{P} \approx \frac{1}{N_{\mathrm{Samp}}} \sum_{i=1}^{N_{\mathrm{Samp}}} f(\mathbf{s}_i) .
\label{eq:FCPsample}
\end{align}
one obtains an estimate of the FCP.  By the central limit theorem, the number of samples required to reach a desired precision $\epsilon_{\mathrm{FCP}}$ scales as $\mathrm{Var}(f)/\epsilon_{\mathrm{FCP}}^2$.  As the Kronecker delta function is constrained to have a value of either $0$ or $1$, the variance of $f$ in this case can be bounded by $1$, and the number of expected samples to converge FCP for a given frequency may be bounded by a constant dictated by the fixed precision $\epsilon_{\mathrm{FCP}}$.  We note also that this constant bound is an upper bound on the number of required samples, and some distributions and experiments will require far less samples.  For example, distributions with a small number of peaks (at the resolution determined by $\Delta_\mathrm{vib}$) may converge extremely rapidly.

In Fig.~\ref{fig:formicacid_MC}, we simulate the results of what one would expect, accordingly, in a boson sampling simulation of Formic acid. This is done by stochastic sampling (Eq.~\eqref{eq:FCPsample}) on the known probability distributions for the output modes and analysed according to Eq.~\eqref{eq:FCP}. The results from this simulation indicate relatively few samples are needed from the device to resolve the overall shape of the spectrum, supporting the experimental feasibility of the approach.

\begin{thebibliography}{50}%
\makeatletter
\providecommand \@ifxundefined [1]{%
 \@ifx{#1\undefined}
}%
\providecommand \@ifnum [1]{%
 \ifnum #1\expandafter \@firstoftwo
 \else \expandafter \@secondoftwo
 \fi
}%
\providecommand \@ifx [1]{%
 \ifx #1\expandafter \@firstoftwo
 \else \expandafter \@secondoftwo
 \fi
}%
\providecommand \natexlab [1]{#1}%
\providecommand \enquote  [1]{``#1''}%
\providecommand \bibnamefont  [1]{#1}%
\providecommand \bibfnamefont [1]{#1}%
\providecommand \citenamefont [1]{#1}%
\providecommand \href@noop [0]{\@secondoftwo}%
\providecommand \href [0]{\begingroup \@sanitize@url \@href}%
\providecommand \@href[1]{\@@startlink{#1}\@@href}%
\providecommand \@@href[1]{\endgroup#1\@@endlink}%
\providecommand \@sanitize@url [0]{\catcode `\\12\catcode `\$12\catcode
  `\&12\catcode `\#12\catcode `\^12\catcode `\_12\catcode `\%12\relax}%
\providecommand \@@startlink[1]{}%
\providecommand \@@endlink[0]{}%
\providecommand \url  [0]{\begingroup\@sanitize@url \@url }%
\providecommand \@url [1]{\endgroup\@href {#1}{\urlprefix }}%
\providecommand \urlprefix  [0]{URL }%
\providecommand \Eprint [0]{\href }%
\providecommand \doibase [0]{http://dx.doi.org/}%
\providecommand \selectlanguage [0]{\@gobble}%
\providecommand \bibinfo  [0]{\@secondoftwo}%
\providecommand \bibfield  [0]{\@secondoftwo}%
\providecommand \translation [1]{[#1]}%
\providecommand \BibitemOpen [0]{}%
\providecommand \bibitemStop [0]{}%
\providecommand \bibitemNoStop [0]{.\EOS\space}%
\providecommand \EOS [0]{\spacefactor3000\relax}%
\providecommand \BibitemShut  [1]{\csname bibitem#1\endcsname}%
\let\auto@bib@innerbib\@empty
\bibitem [{\citenamefont {Deutsch}\ and\ \citenamefont
  {Jozsa}(1992)}]{Deutsch1992}%
  \BibitemOpen
  \bibfield  {author} {\bibinfo {author} {\bibfnamefont {D.}~\bibnamefont
  {Deutsch}}\ and\ \bibinfo {author} {\bibfnamefont {R.}~\bibnamefont
  {Jozsa}},\ }\href {\doibase 10.1098/rspa.1992.0167} {\bibfield  {journal}
  {\bibinfo  {journal} {Proc. R. Soc. London, Ser. A}\ }\textbf {\bibinfo
  {volume} {439}},\ \bibinfo {pages} {553} (\bibinfo {year}
  {1992})}\BibitemShut {NoStop}%
\bibitem [{\citenamefont {Grover}(1997)}]{Grover1997}%
  \BibitemOpen
  \bibfield  {author} {\bibinfo {author} {\bibfnamefont {L.~K.}\ \bibnamefont
  {Grover}},\ }\href {http://link.aps.org/doi/10.1103/PhysRevLett.79.325}
  {\bibfield  {journal} {\bibinfo  {journal} {Phys. Rev. Lett.}\ }\textbf
  {\bibinfo {volume} {79}},\ \bibinfo {pages} {325} (\bibinfo {year}
  {1997})}\BibitemShut {NoStop}%
\bibitem [{\citenamefont {Shor}(1999)}]{Shor1999}%
  \BibitemOpen
  \bibfield  {author} {\bibinfo {author} {\bibfnamefont {P.~W.}\ \bibnamefont
  {Shor}},\ }\href {http://www.jstor.org/stable/2653075} {\bibfield  {journal}
  {\bibinfo  {journal} {SIAM Rev.}\ }\textbf {\bibinfo {volume} {41}},\
  \bibinfo {pages} {303} (\bibinfo {year} {1999})}\BibitemShut {NoStop}%
\bibitem [{\citenamefont {Georgescu}\ \emph {et~al.}(2014)\citenamefont
  {Georgescu}, \citenamefont {Ashhab},\ and\ \citenamefont
  {Nori}}]{Georgescu2014}%
  \BibitemOpen
  \bibfield  {author} {\bibinfo {author} {\bibfnamefont {I.~M.}\ \bibnamefont
  {Georgescu}}, \bibinfo {author} {\bibfnamefont {S.}~\bibnamefont {Ashhab}}, \
  and\ \bibinfo {author} {\bibfnamefont {F.}~\bibnamefont {Nori}},\ }\href
  {\doibase 10.1103/RevModPhys.86.153} {\bibfield  {journal} {\bibinfo
  {journal} {Rev. Mod. Phys.}\ }\textbf {\bibinfo {volume} {86}},\ \bibinfo
  {pages} {153} (\bibinfo {year} {2014})}\BibitemShut {NoStop}%
\bibitem [{\citenamefont {Aspuru-Guzik}\ and\ \citenamefont
  {Walther}(2012)}]{Aspuru-Guzik2012}%
  \BibitemOpen
  \bibfield  {author} {\bibinfo {author} {\bibfnamefont {A.}~\bibnamefont
  {Aspuru-Guzik}}\ and\ \bibinfo {author} {\bibfnamefont {P.}~\bibnamefont
  {Walther}},\ }\href {\doibase 10.1038/nphys2253} {\bibfield  {journal}
  {\bibinfo  {journal} {Nature Phys.}\ }\textbf {\bibinfo {volume} {8}},\
  \bibinfo {pages} {285} (\bibinfo {year} {2012})}\BibitemShut {NoStop}%
\bibitem [{\citenamefont {Bloch}\ \emph {et~al.}(2012)\citenamefont {Bloch},
  \citenamefont {Dalibard},\ and\ \citenamefont {Nascimb\`{e}ne}}]{Bloch2012}%
  \BibitemOpen
  \bibfield  {author} {\bibinfo {author} {\bibfnamefont {I.}~\bibnamefont
  {Bloch}}, \bibinfo {author} {\bibfnamefont {J.}~\bibnamefont {Dalibard}}, \
  and\ \bibinfo {author} {\bibfnamefont {S.}~\bibnamefont {Nascimb\`{e}ne}},\
  }\href {\doibase 10.1038/nphys2259} {\bibfield  {journal} {\bibinfo
  {journal} {Nature Phys.}\ }\textbf {\bibinfo {volume} {8}},\ \bibinfo {pages}
  {267} (\bibinfo {year} {2012})}\BibitemShut {NoStop}%
\bibitem [{\citenamefont {Blatt}\ and\ \citenamefont {Roos}(2012)}]{Blatt2012}%
  \BibitemOpen
  \bibfield  {author} {\bibinfo {author} {\bibfnamefont {R.}~\bibnamefont
  {Blatt}}\ and\ \bibinfo {author} {\bibfnamefont {C.~F.}\ \bibnamefont
  {Roos}},\ }\href {\doibase 10.1038/nphys2252} {\bibfield  {journal} {\bibinfo
   {journal} {Nature Phys.}\ }\textbf {\bibinfo {volume} {8}},\ \bibinfo
  {pages} {277} (\bibinfo {year} {2012})}\BibitemShut {NoStop}%
\bibitem [{\citenamefont {Lloyd}(1996)}]{Lloyd1996}%
  \BibitemOpen
  \bibfield  {author} {\bibinfo {author} {\bibfnamefont {S.}~\bibnamefont
  {Lloyd}},\ }\href {http://research.physics.illinois.edu/demarco/lloyd 96
  paper.pdf} {\bibfield  {journal} {\bibinfo  {journal} {Science}\ }\textbf
  {\bibinfo {volume} {273}},\ \bibinfo {pages} {1073} (\bibinfo {year}
  {1996})}\BibitemShut {NoStop}%
\bibitem [{\citenamefont {Aspuru-Guzik}\ \emph {et~al.}(2005)\citenamefont
  {Aspuru-Guzik}, \citenamefont {Dutoi}, \citenamefont {Love},\ and\
  \citenamefont {Head-Gordon}}]{Aspuru-Guzik2005}%
  \BibitemOpen
  \bibfield  {author} {\bibinfo {author} {\bibfnamefont {A.}~\bibnamefont
  {Aspuru-Guzik}}, \bibinfo {author} {\bibfnamefont {A.~D.}\ \bibnamefont
  {Dutoi}}, \bibinfo {author} {\bibfnamefont {P.~J.}\ \bibnamefont {Love}}, \
  and\ \bibinfo {author} {\bibfnamefont {M.}~\bibnamefont {Head-Gordon}},\
  }\href {\doibase 10.1126/science.1113479} {\bibfield  {journal} {\bibinfo
  {journal} {Science}\ }\textbf {\bibinfo {volume} {309}},\ \bibinfo {pages}
  {1704} (\bibinfo {year} {2005})}\BibitemShut {NoStop}%
\bibitem [{\citenamefont {Kassal}\ \emph {et~al.}(2011)\citenamefont {Kassal},
  \citenamefont {Whitfield}, \citenamefont {Perdomo-Ortiz}, \citenamefont
  {Yung},\ and\ \citenamefont {Aspuru-Guzik}}]{Kassal2011}%
  \BibitemOpen
  \bibfield  {author} {\bibinfo {author} {\bibfnamefont {I.}~\bibnamefont
  {Kassal}}, \bibinfo {author} {\bibfnamefont {J.~D.}\ \bibnamefont
  {Whitfield}}, \bibinfo {author} {\bibfnamefont {A.}~\bibnamefont
  {Perdomo-Ortiz}}, \bibinfo {author} {\bibfnamefont {M.-H.}\ \bibnamefont
  {Yung}}, \ and\ \bibinfo {author} {\bibfnamefont {A.}~\bibnamefont
  {Aspuru-Guzik}},\ }\href {\doibase 10.1146/annurev-physchem-032210-103512}
  {\bibfield  {journal} {\bibinfo  {journal} {Ann. Rev. Phys. Chem.}\ }\textbf
  {\bibinfo {volume} {62}},\ \bibinfo {pages} {185} (\bibinfo {year}
  {2011})}\BibitemShut {NoStop}%
\bibitem [{\citenamefont {Babbush}\ \emph {et~al.}(2014)\citenamefont
  {Babbush}, \citenamefont {McClean}, \citenamefont {Wecker}, \citenamefont
  {Al\'{a}n},\ and\ \citenamefont {Wiebe}}]{Babbush2014}%
  \BibitemOpen
  \bibfield  {author} {\bibinfo {author} {\bibfnamefont {R.}~\bibnamefont
  {Babbush}}, \bibinfo {author} {\bibfnamefont {J.}~\bibnamefont {McClean}},
  \bibinfo {author} {\bibfnamefont {D.}~\bibnamefont {Wecker}}, \bibinfo
  {author} {\bibnamefont {Al\'{a}n}}, \ and\ \bibinfo {author} {\bibfnamefont
  {N.}~\bibnamefont {Wiebe}},\ }\href@noop {} {\bibfield  {journal} {\bibinfo
  {journal} {arXiv: 1410.8159v1}\ } (\bibinfo {year} {2014})}\BibitemShut
  {NoStop}%
\bibitem [{\citenamefont {Aaronson}\ and\ \citenamefont
  {Arkhipov}(2011)}]{Aaronson2011}%
  \BibitemOpen
  \bibfield  {author} {\bibinfo {author} {\bibfnamefont {S.}~\bibnamefont
  {Aaronson}}\ and\ \bibinfo {author} {\bibfnamefont {A.}~\bibnamefont
  {Arkhipov}},\ }\href {\doibase 10.1145/1993636.1993682} {\bibfield  {journal}
  {\bibinfo  {journal} {Proceedings of the 43rd annual ACM symposium on Theory
  of computing - STOC '11}\ ,\ \bibinfo {pages} {333}} (\bibinfo {year}
  {2011})}\BibitemShut {NoStop}%
\bibitem [{\citenamefont {Spring}\ \emph {et~al.}(2013)\citenamefont {Spring},
  \citenamefont {Metcalf}, \citenamefont {Humphreys}, \citenamefont
  {Kolthammer}, \citenamefont {Jin}, \citenamefont {Barbieri}, \citenamefont
  {Datta}, \citenamefont {Thomas-Peter}, \citenamefont {Langford},
  \citenamefont {Kundys}, \citenamefont {Gates}, \citenamefont {Smith},
  \citenamefont {Smith},\ and\ \citenamefont {Walmsley}}]{Spring2013}%
  \BibitemOpen
  \bibfield  {author} {\bibinfo {author} {\bibfnamefont {J.~B.}\ \bibnamefont
  {Spring}}, \bibinfo {author} {\bibfnamefont {B.~J.}\ \bibnamefont {Metcalf}},
  \bibinfo {author} {\bibfnamefont {P.~C.}\ \bibnamefont {Humphreys}}, \bibinfo
  {author} {\bibfnamefont {W.~S.}\ \bibnamefont {Kolthammer}}, \bibinfo
  {author} {\bibfnamefont {X.-M.}\ \bibnamefont {Jin}}, \bibinfo {author}
  {\bibfnamefont {M.}~\bibnamefont {Barbieri}}, \bibinfo {author}
  {\bibfnamefont {A.}~\bibnamefont {Datta}}, \bibinfo {author} {\bibfnamefont
  {N.}~\bibnamefont {Thomas-Peter}}, \bibinfo {author} {\bibfnamefont {N.~K.}\
  \bibnamefont {Langford}}, \bibinfo {author} {\bibfnamefont {D.}~\bibnamefont
  {Kundys}}, \bibinfo {author} {\bibfnamefont {J.~C.}\ \bibnamefont {Gates}},
  \bibinfo {author} {\bibfnamefont {B.~J.}\ \bibnamefont {Smith}}, \bibinfo
  {author} {\bibfnamefont {P.~G.~R.}\ \bibnamefont {Smith}}, \ and\ \bibinfo
  {author} {\bibfnamefont {I.~A.}\ \bibnamefont {Walmsley}},\ }\href {\doibase
  10.1126/science.1231692} {\bibfield  {journal} {\bibinfo  {journal}
  {Science}\ }\textbf {\bibinfo {volume} {339}},\ \bibinfo {pages} {798}
  (\bibinfo {year} {2013})}\BibitemShut {NoStop}%
\bibitem [{\citenamefont {Broome}\ \emph {et~al.}(2013)\citenamefont {Broome},
  \citenamefont {Fedrizzi}, \citenamefont {Rahimi-Keshari}, \citenamefont
  {Dove}, \citenamefont {Aaronson}, \citenamefont {Ralph},\ and\ \citenamefont
  {White}}]{Broome2013}%
  \BibitemOpen
  \bibfield  {author} {\bibinfo {author} {\bibfnamefont {M.~A.}\ \bibnamefont
  {Broome}}, \bibinfo {author} {\bibfnamefont {A.}~\bibnamefont {Fedrizzi}},
  \bibinfo {author} {\bibfnamefont {S.}~\bibnamefont {Rahimi-Keshari}},
  \bibinfo {author} {\bibfnamefont {J.}~\bibnamefont {Dove}}, \bibinfo {author}
  {\bibfnamefont {S.}~\bibnamefont {Aaronson}}, \bibinfo {author}
  {\bibfnamefont {T.~C.}\ \bibnamefont {Ralph}}, \ and\ \bibinfo {author}
  {\bibfnamefont {A.~G.}\ \bibnamefont {White}},\ }\href {\doibase
  10.1126/science.1231440} {\bibfield  {journal} {\bibinfo  {journal}
  {Science}\ }\textbf {\bibinfo {volume} {339}},\ \bibinfo {pages} {794}
  (\bibinfo {year} {2013})}\BibitemShut {NoStop}%
\bibitem [{\citenamefont {Crespi}\ \emph {et~al.}(2013)\citenamefont {Crespi},
  \citenamefont {Osellame}, \citenamefont {Ramponi}, \citenamefont {Brod},
  \citenamefont {Galvao}, \citenamefont {Spagnolo}, \citenamefont {Vitelli},
  \citenamefont {Maiorino}, \citenamefont {Mataloni},\ and\ \citenamefont
  {Sciarrino}}]{Crespi2013}%
  \BibitemOpen
  \bibfield  {author} {\bibinfo {author} {\bibfnamefont {A.}~\bibnamefont
  {Crespi}}, \bibinfo {author} {\bibfnamefont {R.}~\bibnamefont {Osellame}},
  \bibinfo {author} {\bibfnamefont {R.}~\bibnamefont {Ramponi}}, \bibinfo
  {author} {\bibfnamefont {D.~J.}\ \bibnamefont {Brod}}, \bibinfo {author}
  {\bibfnamefont {E.~F.}\ \bibnamefont {Galvao}}, \bibinfo {author}
  {\bibfnamefont {N.}~\bibnamefont {Spagnolo}}, \bibinfo {author}
  {\bibfnamefont {C.}~\bibnamefont {Vitelli}}, \bibinfo {author} {\bibfnamefont
  {E.}~\bibnamefont {Maiorino}}, \bibinfo {author} {\bibfnamefont
  {P.}~\bibnamefont {Mataloni}}, \ and\ \bibinfo {author} {\bibfnamefont
  {F.}~\bibnamefont {Sciarrino}},\ }\href {\doibase 10.1038/NPHOTON.2013.112}
  {\bibfield  {journal} {\bibinfo  {journal} {Nature Photon.}\ }\textbf
  {\bibinfo {volume} {7}},\ \bibinfo {pages} {545} (\bibinfo {year}
  {2013})}\BibitemShut {NoStop}%
\bibitem [{\citenamefont {Tillmann}\ \emph {et~al.}(2013)\citenamefont
  {Tillmann}, \citenamefont {Dakic}, \citenamefont {Heilmann}, \citenamefont
  {Nolte}, \citenamefont {Szameit},\ and\ \citenamefont
  {Walther}}]{Tillmann2013}%
  \BibitemOpen
  \bibfield  {author} {\bibinfo {author} {\bibfnamefont {M.}~\bibnamefont
  {Tillmann}}, \bibinfo {author} {\bibfnamefont {B.}~\bibnamefont {Dakic}},
  \bibinfo {author} {\bibfnamefont {R.}~\bibnamefont {Heilmann}}, \bibinfo
  {author} {\bibfnamefont {S.}~\bibnamefont {Nolte}}, \bibinfo {author}
  {\bibfnamefont {A.}~\bibnamefont {Szameit}}, \ and\ \bibinfo {author}
  {\bibfnamefont {P.}~\bibnamefont {Walther}},\ }\href {\doibase
  10.1038/NPHOTON.2013.102} {\bibfield  {journal} {\bibinfo  {journal} {Nature
  Photon.}\ }\textbf {\bibinfo {volume} {7}},\ \bibinfo {pages} {540} (\bibinfo
  {year} {2013})}\BibitemShut {NoStop}%
\bibitem [{\citenamefont {Shchesnovich}(2014)}]{Shchesnovich2014}%
  \BibitemOpen
  \bibfield  {author} {\bibinfo {author} {\bibfnamefont {V.~S.}\ \bibnamefont
  {Shchesnovich}},\ }\href@noop {} {\bibfield  {journal} {\bibinfo  {journal}
  {arXiv: 1403.4459v6}\ } (\bibinfo {year} {2014})}\BibitemShut {NoStop}%
\bibitem [{\citenamefont {Rohde}\ \emph {et~al.}(2014)\citenamefont {Rohde},
  \citenamefont {Motes}, \citenamefont {Knott},\ and\ \citenamefont
  {Munro}}]{Rohde2014}%
  \BibitemOpen
  \bibfield  {author} {\bibinfo {author} {\bibfnamefont {P.~P.}\ \bibnamefont
  {Rohde}}, \bibinfo {author} {\bibfnamefont {K.~R.}\ \bibnamefont {Motes}},
  \bibinfo {author} {\bibfnamefont {P.~A.}\ \bibnamefont {Knott}}, \ and\
  \bibinfo {author} {\bibfnamefont {W.~J.}\ \bibnamefont {Munro}},\ }\href@noop
  {} {\bibfield  {journal} {\bibinfo  {journal} {arXiv: 1401.2199v2}\ }
  (\bibinfo {year} {2014})}\BibitemShut {NoStop}%
\bibitem [{\citenamefont {Sharp}\ and\ \citenamefont
  {Rosenstock}(1964)}]{sharp:1964}%
  \BibitemOpen
  \bibfield  {author} {\bibinfo {author} {\bibfnamefont {T.~E.}\ \bibnamefont
  {Sharp}}\ and\ \bibinfo {author} {\bibfnamefont {H.~M.}\ \bibnamefont
  {Rosenstock}},\ }\href@noop {} {\bibfield  {journal} {\bibinfo  {journal} {J.
  Chem. Phys.}\ }\textbf {\bibinfo {volume} {41}},\ \bibinfo {pages} {3453}
  (\bibinfo {year} {1964})}\BibitemShut {NoStop}%
\bibitem [{\citenamefont {Doktorov}\ \emph {et~al.}(1977)\citenamefont
  {Doktorov}, \citenamefont {Malkin},\ and\ \citenamefont
  {Man'ko}}]{doktorov:1977}%
  \BibitemOpen
  \bibfield  {author} {\bibinfo {author} {\bibfnamefont {E.~V.}\ \bibnamefont
  {Doktorov}}, \bibinfo {author} {\bibfnamefont {I.~A.}\ \bibnamefont
  {Malkin}}, \ and\ \bibinfo {author} {\bibfnamefont {V.~I.}\ \bibnamefont
  {Man'ko}},\ }\href@noop {} {\bibfield  {journal} {\bibinfo  {journal} {J.
  Mol. Spectrosc.}\ }\textbf {\bibinfo {volume} {64}},\ \bibinfo {pages} {302}
  (\bibinfo {year} {1977})}\BibitemShut {NoStop}%
\bibitem [{\citenamefont {Malmqvist}\ and\ \citenamefont
  {Forsberg}(1998)}]{malmqvist:1998}%
  \BibitemOpen
  \bibfield  {author} {\bibinfo {author} {\bibfnamefont {P.-{\AA}.}\
  \bibnamefont {Malmqvist}}\ and\ \bibinfo {author} {\bibfnamefont
  {N.}~\bibnamefont {Forsberg}},\ }\href@noop {} {\bibfield  {journal}
  {\bibinfo  {journal} {Chem. Phys.}\ }\textbf {\bibinfo {volume} {228}},\
  \bibinfo {pages} {227} (\bibinfo {year} {1998})}\BibitemShut {NoStop}%
\bibitem [{\citenamefont {Ruhoff}\ and\ \citenamefont
  {Ratner}(2000)}]{cjruhoff:2000}%
  \BibitemOpen
  \bibfield  {author} {\bibinfo {author} {\bibfnamefont {P.~T.}\ \bibnamefont
  {Ruhoff}}\ and\ \bibinfo {author} {\bibfnamefont {M.~A.}\ \bibnamefont
  {Ratner}},\ }\href@noop {} {\bibfield  {journal} {\bibinfo  {journal} {Inter.
  J. Quantum Chem.}\ }\textbf {\bibinfo {volume} {77}},\ \bibinfo {pages} {383}
  (\bibinfo {year} {2000})}\BibitemShut {NoStop}%
\bibitem [{\citenamefont {Jankowiak}\ \emph {et~al.}(2007)\citenamefont
  {Jankowiak}, \citenamefont {Stuber},\ and\ \citenamefont
  {Berger}}]{jankowiak:2007}%
  \BibitemOpen
  \bibfield  {author} {\bibinfo {author} {\bibfnamefont {H.-C.}\ \bibnamefont
  {Jankowiak}}, \bibinfo {author} {\bibfnamefont {J.~L.}\ \bibnamefont
  {Stuber}}, \ and\ \bibinfo {author} {\bibfnamefont {R.}~\bibnamefont
  {Berger}},\ }\href@noop {} {\bibfield  {journal} {\bibinfo  {journal} {J.
  Chem. Phys.}\ }\textbf {\bibinfo {volume} {127}},\ \bibinfo {pages} {234101}
  (\bibinfo {year} {2007})}\BibitemShut {NoStop}%
\bibitem [{\citenamefont {Santoro}\ \emph {et~al.}(2007)\citenamefont
  {Santoro}, \citenamefont {Lami}, \citenamefont {Improta},\ and\ \citenamefont
  {Barone}}]{santoro:2007b}%
  \BibitemOpen
  \bibfield  {author} {\bibinfo {author} {\bibfnamefont {F.}~\bibnamefont
  {Santoro}}, \bibinfo {author} {\bibfnamefont {A.}~\bibnamefont {Lami}},
  \bibinfo {author} {\bibfnamefont {R.}~\bibnamefont {Improta}}, \ and\
  \bibinfo {author} {\bibfnamefont {V.}~\bibnamefont {Barone}},\ }\href@noop {}
  {\bibfield  {journal} {\bibinfo  {journal} {J. Chem. Phys.}\ }\textbf
  {\bibinfo {volume} {126}} (\bibinfo {year} {2007})}\BibitemShut {NoStop}%
\bibitem [{\citenamefont {Duschinsky}(1937)}]{duschinsky:1937}%
  \BibitemOpen
  \bibfield  {author} {\bibinfo {author} {\bibfnamefont {F.}~\bibnamefont
  {Duschinsky}},\ }\href@noop {} {\bibfield  {journal} {\bibinfo  {journal}
  {Acta Physicochim. URSS}\ }\textbf {\bibinfo {volume} {7}},\ \bibinfo {pages}
  {551} (\bibinfo {year} {1937})}\BibitemShut {NoStop}%
\bibitem [{\citenamefont {Mart\'{i}n-L\'opez}\ \emph
  {et~al.}(tted)\citenamefont {Mart\'{i}n-L\'opez}, \citenamefont {Russell},
  \citenamefont {Sparrow}, \citenamefont {O'Brien},\ and\ \citenamefont
  {Laing}}]{lopez2014}%
  \BibitemOpen
  \bibfield  {author} {\bibinfo {author} {\bibfnamefont {E.}~\bibnamefont
  {Mart\'{i}n-L\'opez}}, \bibinfo {author} {\bibfnamefont {N.~J.}\ \bibnamefont
  {Russell}}, \bibinfo {author} {\bibfnamefont {C.}~\bibnamefont {Sparrow}},
  \bibinfo {author} {\bibfnamefont {J.~L.}\ \bibnamefont {O'Brien}}, \ and\
  \bibinfo {author} {\bibfnamefont {A.}~\bibnamefont {Laing}},\ }\href@noop {}
  {\  (\bibinfo {year} {To be submitted})}\BibitemShut {NoStop}%
\bibitem [{\citenamefont {Ma}\ and\ \citenamefont {Phodes}(1990)}]{Ma1990}%
  \BibitemOpen
  \bibfield  {author} {\bibinfo {author} {\bibfnamefont {X.}~\bibnamefont
  {Ma}}\ and\ \bibinfo {author} {\bibfnamefont {W.}~\bibnamefont {Phodes}},\
  }\href@noop {} {\bibfield  {journal} {\bibinfo  {journal} {Phys. Rev. A}\
  }\textbf {\bibinfo {volume} {41}},\ \bibinfo {pages} {4625} (\bibinfo {year}
  {1990})}\BibitemShut {NoStop}%
\bibitem [{\citenamefont {Scheel}(2004)}]{Scheel2004}%
  \BibitemOpen
  \bibfield  {author} {\bibinfo {author} {\bibfnamefont {S.}~\bibnamefont
  {Scheel}},\ }\href@noop {} {\bibfield  {journal} {\bibinfo  {journal}
  {arXiv:quant-ph/0406127}\ } (\bibinfo {year} {2004})}\BibitemShut {NoStop}%
\bibitem [{\citenamefont {Hachmann}\ \emph {et~al.}(2011)\citenamefont
  {Hachmann}, \citenamefont {Olivares-Amaya}, \citenamefont {Atahan-Evrenk},
  \citenamefont {Amador-Bedolla}, \citenamefont {Sánchez-Carrera},
  \citenamefont {Gold-Parker}, \citenamefont {Vogt}, \citenamefont {Brockway},\
  and\ \citenamefont {Aspuru-Guzik}}]{Johannes2011}%
  \BibitemOpen
  \bibfield  {author} {\bibinfo {author} {\bibfnamefont {J.}~\bibnamefont
  {Hachmann}}, \bibinfo {author} {\bibfnamefont {R.}~\bibnamefont
  {Olivares-Amaya}}, \bibinfo {author} {\bibfnamefont {S.}~\bibnamefont
  {Atahan-Evrenk}}, \bibinfo {author} {\bibfnamefont {C.}~\bibnamefont
  {Amador-Bedolla}}, \bibinfo {author} {\bibfnamefont {R.~S.}\ \bibnamefont
  {Sánchez-Carrera}}, \bibinfo {author} {\bibfnamefont {A.}~\bibnamefont
  {Gold-Parker}}, \bibinfo {author} {\bibfnamefont {L.}~\bibnamefont {Vogt}},
  \bibinfo {author} {\bibfnamefont {A.~M.}\ \bibnamefont {Brockway}}, \ and\
  \bibinfo {author} {\bibfnamefont {A.}~\bibnamefont {Aspuru-Guzik}},\
  }\href@noop {} {\bibfield  {journal} {\bibinfo  {journal} {J. Phys. Chem.
  Lett.}\ }\textbf {\bibinfo {volume} {2}},\ \bibinfo {pages} {2241} (\bibinfo
  {year} {2011})}\BibitemShut {NoStop}%
\bibitem [{\citenamefont {Gross}\ \emph {et~al.}(2000)\citenamefont {Gross},
  \citenamefont {M\"uller}, \citenamefont {Nothofer}, \citenamefont {Scherf},
  \citenamefont {Neher}, \citenamefont {Br\"auchle},\ and\ \citenamefont
  {Meerholz}}]{Gross2000}%
  \BibitemOpen
  \bibfield  {author} {\bibinfo {author} {\bibfnamefont {M.}~\bibnamefont
  {Gross}}, \bibinfo {author} {\bibfnamefont {D.~C.}\ \bibnamefont {M\"uller}},
  \bibinfo {author} {\bibfnamefont {H.-G.}\ \bibnamefont {Nothofer}}, \bibinfo
  {author} {\bibfnamefont {U.}~\bibnamefont {Scherf}}, \bibinfo {author}
  {\bibfnamefont {D.}~\bibnamefont {Neher}}, \bibinfo {author} {\bibfnamefont
  {C.}~\bibnamefont {Br\"auchle}}, \ and\ \bibinfo {author} {\bibfnamefont
  {K.}~\bibnamefont {Meerholz}},\ }\href@noop {} {\bibfield  {journal}
  {\bibinfo  {journal} {Nature}\ }\textbf {\bibinfo {volume} {405}},\ \bibinfo
  {pages} {661} (\bibinfo {year} {2000})}\BibitemShut {NoStop}%
\bibitem [{\citenamefont {Winkler}(2013)}]{Winkler29032013}%
  \BibitemOpen
  \bibfield  {author} {\bibinfo {author} {\bibfnamefont {J.~R.}\ \bibnamefont
  {Winkler}},\ }\href@noop {} {\bibfield  {journal} {\bibinfo  {journal}
  {Science}\ }\textbf {\bibinfo {volume} {339}},\ \bibinfo {pages} {1530}
  (\bibinfo {year} {2013})}\BibitemShut {NoStop}%
\bibitem [{\citenamefont {Dierksen}\ and\ \citenamefont
  {Grimme}(2004)}]{Dierksen2004}%
  \BibitemOpen
  \bibfield  {author} {\bibinfo {author} {\bibfnamefont {M.}~\bibnamefont
  {Dierksen}}\ and\ \bibinfo {author} {\bibfnamefont {S.}~\bibnamefont
  {Grimme}},\ }\href@noop {} {\bibfield  {journal} {\bibinfo  {journal} {J.
  Phys. Chem. A}\ }\textbf {\bibinfo {volume} {108}},\ \bibinfo {pages} {10225}
  (\bibinfo {year} {2004})}\BibitemShut {NoStop}%
\bibitem [{\citenamefont {Hayes}\ \emph {et~al.}(2011)\citenamefont {Hayes},
  \citenamefont {Wen}, \citenamefont {Panitchayangkoon}, \citenamefont
  {Blankenship},\ and\ \citenamefont {Engel}}]{Hayes2011}%
  \BibitemOpen
  \bibfield  {author} {\bibinfo {author} {\bibfnamefont {D.}~\bibnamefont
  {Hayes}}, \bibinfo {author} {\bibfnamefont {J.}~\bibnamefont {Wen}}, \bibinfo
  {author} {\bibfnamefont {G.}~\bibnamefont {Panitchayangkoon}}, \bibinfo
  {author} {\bibfnamefont {R.~E.}\ \bibnamefont {Blankenship}}, \ and\ \bibinfo
  {author} {\bibfnamefont {G.~S.}\ \bibnamefont {Engel}},\ }\href@noop {}
  {\bibfield  {journal} {\bibinfo  {journal} {Faraday Discuss.}\ }\textbf
  {\bibinfo {volume} {150}},\ \bibinfo {pages} {459} (\bibinfo {year}
  {2011})}\BibitemShut {NoStop}%
\bibitem [{\citenamefont {Choi}\ \emph {et~al.}(2005)\citenamefont {Choi},
  \citenamefont {Lee},\ and\ \citenamefont {Kim}}]{choi:2005}%
  \BibitemOpen
  \bibfield  {author} {\bibinfo {author} {\bibfnamefont {K.-W.}\ \bibnamefont
  {Choi}}, \bibinfo {author} {\bibfnamefont {J.-H.}\ \bibnamefont {Lee}}, \
  and\ \bibinfo {author} {\bibfnamefont {S.~K.}\ \bibnamefont {Kim}},\
  }\href@noop {} {\bibfield  {journal} {\bibinfo  {journal} {J. Am. Chem.
  Soc.}\ }\textbf {\bibinfo {volume} {127}},\ \bibinfo {pages} {15674}
  (\bibinfo {year} {2005})}\BibitemShut {NoStop}%
\bibitem [{\citenamefont {Huh}(2011)}]{Huh2011a}%
  \BibitemOpen
  \bibfield  {author} {\bibinfo {author} {\bibfnamefont {J.}~\bibnamefont
  {Huh}},\ }\emph {\bibinfo {title} {{Unified description of vibronic
  transitions with coherent states}}},\ \href@noop {} {Ph.D. thesis} (\bibinfo
  {year} {2011}),\ \Eprint
  {http://arxiv.org/abs/http://publikationen.stub.uni-frankfurt.de/frontdoor/index/index/docId/21033}
  {http://publikationen.stub.uni-frankfurt.de/frontdoor/index/index/docId/21033}
  \BibitemShut {NoStop}%
\bibitem [{\citenamefont {Huh}\ and\ \citenamefont {Berger}(2011)}]{Huh2011}%
  \BibitemOpen
  \bibfield  {author} {\bibinfo {author} {\bibfnamefont {J.}~\bibnamefont
  {Huh}}\ and\ \bibinfo {author} {\bibfnamefont {R.}~\bibnamefont {Berger}},\
  }\href@noop {} {\bibfield  {journal} {\bibinfo  {journal} {Faraday Discuss.}\
  }\textbf {\bibinfo {volume} {150}},\ \bibinfo {pages} {363} (\bibinfo {year}
  {2011})}\BibitemShut {NoStop}%
\bibitem [{\citenamefont {Huh}\ and\ \citenamefont {Berger}(2012)}]{Huh2012}%
  \BibitemOpen
  \bibfield  {author} {\bibinfo {author} {\bibfnamefont {J.}~\bibnamefont
  {Huh}}\ and\ \bibinfo {author} {\bibfnamefont {R.}~\bibnamefont {Berger}},\
  }\href@noop {} {\bibfield  {journal} {\bibinfo  {journal} {J. Phys. Conf.
  Ser.}\ }\textbf {\bibinfo {volume} {380}},\ \bibinfo {pages} {012019}
  (\bibinfo {year} {2012})}\BibitemShut {NoStop}%
\bibitem [{\citenamefont {Kan}(2008)}]{kan:2007}%
  \BibitemOpen
  \bibfield  {author} {\bibinfo {author} {\bibfnamefont {R.}~\bibnamefont
  {Kan}},\ }\href@noop {} {\bibfield  {journal} {\bibinfo  {journal} {J.
  Multivar. Anal.}\ }\textbf {\bibinfo {volume} {99}},\ \bibinfo {pages} {542}
  (\bibinfo {year} {2008})}\BibitemShut {NoStop}%
\bibitem [{\citenamefont {Rahimi-Keshari}\ \emph {et~al.}(2014)\citenamefont
  {Rahimi-Keshari}, \citenamefont {Lund},\ and\ \citenamefont
  {Ralph}}]{rahimi2014}%
  \BibitemOpen
  \bibfield  {author} {\bibinfo {author} {\bibfnamefont {S.}~\bibnamefont
  {Rahimi-Keshari}}, \bibinfo {author} {\bibfnamefont {A.~P.}\ \bibnamefont
  {Lund}}, \ and\ \bibinfo {author} {\bibfnamefont {T.~C.}\ \bibnamefont
  {Ralph}},\ }\href@noop {} {\bibfield  {journal} {\bibinfo  {journal} {arXiv:
  1408.3712v1}\ } (\bibinfo {year} {2014})}\BibitemShut {NoStop}%
\bibitem [{\citenamefont {Berger}\ and\ \citenamefont
  {Klessinger}(1997)}]{berger:1997}%
  \BibitemOpen
  \bibfield  {author} {\bibinfo {author} {\bibfnamefont {R.}~\bibnamefont
  {Berger}}\ and\ \bibinfo {author} {\bibfnamefont {M.}~\bibnamefont
  {Klessinger}},\ }\href@noop {} {\bibfield  {journal} {\bibinfo  {journal} {J.
  Comput. Chem.}\ }\textbf {\bibinfo {volume} {18}},\ \bibinfo {pages} {1312}
  (\bibinfo {year} {1997})}\BibitemShut {NoStop}%
\bibitem [{\citenamefont {Lund}\ \emph {et~al.}(2014)\citenamefont {Lund},
  \citenamefont {Laing}, \citenamefont {Rahimi-Keshari}, \citenamefont
  {Rudolph}, \citenamefont {O'Brien},\ and\ \citenamefont
  {Ralph}}]{PhysRevLett.113.100502}%
  \BibitemOpen
  \bibfield  {author} {\bibinfo {author} {\bibfnamefont {A.~P.}\ \bibnamefont
  {Lund}}, \bibinfo {author} {\bibfnamefont {A.}~\bibnamefont {Laing}},
  \bibinfo {author} {\bibfnamefont {S.}~\bibnamefont {Rahimi-Keshari}},
  \bibinfo {author} {\bibfnamefont {T.}~\bibnamefont {Rudolph}}, \bibinfo
  {author} {\bibfnamefont {J.~L.}\ \bibnamefont {O'Brien}}, \ and\ \bibinfo
  {author} {\bibfnamefont {T.~C.}\ \bibnamefont {Ralph}},\ }\href {\doibase
  10.1103/PhysRevLett.113.100502} {\bibfield  {journal} {\bibinfo  {journal}
  {Phys. Rev. Lett.}\ }\textbf {\bibinfo {volume} {113}},\ \bibinfo {pages}
  {100502} (\bibinfo {year} {2014})}\BibitemShut {NoStop}%
\bibitem [{\citenamefont {Santoro}\ \emph {et~al.}(2008)\citenamefont
  {Santoro}, \citenamefont {Lami}, \citenamefont {Improta}, \citenamefont
  {Bloino},\ and\ \citenamefont {Barone}}]{santoro:2008}%
  \BibitemOpen
  \bibfield  {author} {\bibinfo {author} {\bibfnamefont {F.}~\bibnamefont
  {Santoro}}, \bibinfo {author} {\bibfnamefont {A.}~\bibnamefont {Lami}},
  \bibinfo {author} {\bibfnamefont {R.}~\bibnamefont {Improta}}, \bibinfo
  {author} {\bibfnamefont {J.}~\bibnamefont {Bloino}}, \ and\ \bibinfo {author}
  {\bibfnamefont {V.}~\bibnamefont {Barone}},\ }\href@noop {} {\bibfield
  {journal} {\bibinfo  {journal} {J. Chem. Phys.}\ }\textbf {\bibinfo {volume}
  {128}},\ \bibinfo {pages} {224311} (\bibinfo {year} {2008})}\BibitemShut
  {NoStop}%
\bibitem [{\citenamefont {Olson}\ \emph {et~al.}(2014)\citenamefont {Olson},
  \citenamefont {Seshadreesan}, \citenamefont {Motes}, \citenamefont {Rohde},\
  and\ \citenamefont {Dowling}}]{olson2014}%
  \BibitemOpen
  \bibfield  {author} {\bibinfo {author} {\bibfnamefont {J.~P.}\ \bibnamefont
  {Olson}}, \bibinfo {author} {\bibfnamefont {K.~P.}\ \bibnamefont
  {Seshadreesan}}, \bibinfo {author} {\bibfnamefont {K.~R.}\ \bibnamefont
  {Motes}}, \bibinfo {author} {\bibfnamefont {P.~P.}\ \bibnamefont {Rohde}}, \
  and\ \bibinfo {author} {\bibfnamefont {J.~P.}\ \bibnamefont {Dowling}},\
  }\href@noop {} {\bibfield  {journal} {\bibinfo  {journal} {arXiv:
  1406.7821v2}\ } (\bibinfo {year} {2014})}\BibitemShut {NoStop}%
\bibitem [{\citenamefont {Leach}\ \emph {et~al.}(2003)\citenamefont {Leach},
  \citenamefont {Schwell}, \citenamefont {Talbi}, \citenamefont {Berthier},
  \citenamefont {Hottmann}, \citenamefont {Jochims},\ and\ \citenamefont
  {Baumg\"artel}}]{leach:2003}%
  \BibitemOpen
  \bibfield  {author} {\bibinfo {author} {\bibfnamefont {S.}~\bibnamefont
  {Leach}}, \bibinfo {author} {\bibfnamefont {M.}~\bibnamefont {Schwell}},
  \bibinfo {author} {\bibfnamefont {D.}~\bibnamefont {Talbi}}, \bibinfo
  {author} {\bibfnamefont {G.}~\bibnamefont {Berthier}}, \bibinfo {author}
  {\bibfnamefont {K.}~\bibnamefont {Hottmann}}, \bibinfo {author}
  {\bibfnamefont {H.-W.}\ \bibnamefont {Jochims}}, \ and\ \bibinfo {author}
  {\bibfnamefont {H.}~\bibnamefont {Baumg\"artel}},\ }\href@noop {} {\bibfield
  {journal} {\bibinfo  {journal} {Chem. Phys.}\ }\textbf {\bibinfo {volume}
  {286}},\ \bibinfo {pages} {15} (\bibinfo {year} {2003})}\BibitemShut
  {NoStop}%
\bibitem [{\citenamefont {Berger}\ \emph {et~al.}(1998)\citenamefont {Berger},
  \citenamefont {Fischer},\ and\ \citenamefont {Klessinger}}]{berger:1998}%
  \BibitemOpen
  \bibfield  {author} {\bibinfo {author} {\bibfnamefont {R.}~\bibnamefont
  {Berger}}, \bibinfo {author} {\bibfnamefont {C.}~\bibnamefont {Fischer}}, \
  and\ \bibinfo {author} {\bibfnamefont {M.}~\bibnamefont {Klessinger}},\
  }\href@noop {} {\bibfield  {journal} {\bibinfo  {journal} {J. Phys. Chem.}\
  }\textbf {\bibinfo {volume} {102}},\ \bibinfo {pages} {7157} (\bibinfo {year}
  {1998})}\BibitemShut {NoStop}%
\bibitem [{\citenamefont {Carolan}\ \emph {et~al.}(2014)\citenamefont
  {Carolan}, \citenamefont {Meinecke}, \citenamefont {Shadbolt}, \citenamefont
  {Russell}, \citenamefont {Ismail}, \citenamefont {W\"{o}rhoff}, \citenamefont
  {Rudolph}, \citenamefont {Thompson}, \citenamefont {O'Brien}, \citenamefont
  {Matthews},\ and\ \citenamefont {Laing}}]{Carolan2014}%
  \BibitemOpen
  \bibfield  {author} {\bibinfo {author} {\bibfnamefont {J.}~\bibnamefont
  {Carolan}}, \bibinfo {author} {\bibfnamefont {J.~D.~a.}\ \bibnamefont
  {Meinecke}}, \bibinfo {author} {\bibfnamefont {P.~J.}\ \bibnamefont
  {Shadbolt}}, \bibinfo {author} {\bibfnamefont {N.~J.}\ \bibnamefont
  {Russell}}, \bibinfo {author} {\bibfnamefont {N.}~\bibnamefont {Ismail}},
  \bibinfo {author} {\bibfnamefont {K.}~\bibnamefont {W\"{o}rhoff}}, \bibinfo
  {author} {\bibfnamefont {T.}~\bibnamefont {Rudolph}}, \bibinfo {author}
  {\bibfnamefont {M.~G.}\ \bibnamefont {Thompson}}, \bibinfo {author}
  {\bibfnamefont {J.~L.}\ \bibnamefont {O'Brien}}, \bibinfo {author}
  {\bibfnamefont {J.~C.~F.}\ \bibnamefont {Matthews}}, \ and\ \bibinfo {author}
  {\bibfnamefont {A.}~\bibnamefont {Laing}},\ }\href {\doibase
  10.1038/nphoton.2014.152} {\bibfield  {journal} {\bibinfo  {journal} {Nature
  Photon.}\ }\textbf {\bibinfo {volume} {8}},\ \bibinfo {pages} {621} (\bibinfo
  {year} {2014})}\BibitemShut {NoStop}%
\bibitem [{\citenamefont {Josse}\ \emph {et~al.}(2006)\citenamefont {Josse},
  \citenamefont {Sabuncu}, \citenamefont {Cerf}, \citenamefont {Leuchs},\ and\
  \citenamefont {Andersen}}]{Josse2006}%
  \BibitemOpen
  \bibfield  {author} {\bibinfo {author} {\bibfnamefont {V.}~\bibnamefont
  {Josse}}, \bibinfo {author} {\bibfnamefont {M.}~\bibnamefont {Sabuncu}},
  \bibinfo {author} {\bibfnamefont {N.}~\bibnamefont {Cerf}}, \bibinfo {author}
  {\bibfnamefont {G.}~\bibnamefont {Leuchs}}, \ and\ \bibinfo {author}
  {\bibfnamefont {U.}~\bibnamefont {Andersen}},\ }\href {\doibase
  10.1103/PhysRevLett.96.163602} {\bibfield  {journal} {\bibinfo  {journal}
  {Phys. Rev. Lett.}\ }\textbf {\bibinfo {volume} {96}},\ \bibinfo {pages}
  {163602} (\bibinfo {year} {2006})}\BibitemShut {NoStop}%
\bibitem [{\citenamefont {Yoshikawa}\ \emph {et~al.}(2007)\citenamefont
  {Yoshikawa}, \citenamefont {Hayashi}, \citenamefont {Akiyama}, \citenamefont
  {Takei}, \citenamefont {Huck}, \citenamefont {Andersen},\ and\ \citenamefont
  {Furusawa}}]{Yoshikawa2007}%
  \BibitemOpen
  \bibfield  {author} {\bibinfo {author} {\bibfnamefont {J.-i.}\ \bibnamefont
  {Yoshikawa}}, \bibinfo {author} {\bibfnamefont {T.}~\bibnamefont {Hayashi}},
  \bibinfo {author} {\bibfnamefont {T.}~\bibnamefont {Akiyama}}, \bibinfo
  {author} {\bibfnamefont {N.}~\bibnamefont {Takei}}, \bibinfo {author}
  {\bibfnamefont {A.}~\bibnamefont {Huck}}, \bibinfo {author} {\bibfnamefont
  {U.}~\bibnamefont {Andersen}}, \ and\ \bibinfo {author} {\bibfnamefont
  {A.}~\bibnamefont {Furusawa}},\ }\href {\doibase 10.1103/PhysRevA.76.060301}
  {\bibfield  {journal} {\bibinfo  {journal} {Phys. Rev. A}\ }\textbf {\bibinfo
  {volume} {76}},\ \bibinfo {pages} {060301} (\bibinfo {year}
  {2007})}\BibitemShut {NoStop}%
\bibitem [{\citenamefont {Miwa}\ \emph {et~al.}(2014)\citenamefont {Miwa},
  \citenamefont {Yoshikawa}, \citenamefont {Iwata}, \citenamefont {Endo},
  \citenamefont {Marek}, \citenamefont {Filip}, \citenamefont {van Loock},\
  and\ \citenamefont {Furusawa}}]{Miwa2014}%
  \BibitemOpen
  \bibfield  {author} {\bibinfo {author} {\bibfnamefont {Y.}~\bibnamefont
  {Miwa}}, \bibinfo {author} {\bibfnamefont {J.-i.}\ \bibnamefont {Yoshikawa}},
  \bibinfo {author} {\bibfnamefont {N.}~\bibnamefont {Iwata}}, \bibinfo
  {author} {\bibfnamefont {M.}~\bibnamefont {Endo}}, \bibinfo {author}
  {\bibfnamefont {P.}~\bibnamefont {Marek}}, \bibinfo {author} {\bibfnamefont
  {R.}~\bibnamefont {Filip}}, \bibinfo {author} {\bibfnamefont
  {P.}~\bibnamefont {van Loock}}, \ and\ \bibinfo {author} {\bibfnamefont
  {A.}~\bibnamefont {Furusawa}},\ }\href {\doibase
  10.1103/PhysRevLett.113.013601} {\bibfield  {journal} {\bibinfo  {journal}
  {Phys. Rev. Lett.}\ }\textbf {\bibinfo {volume} {113}},\ \bibinfo {pages}
  {013601} (\bibinfo {year} {2014})}\BibitemShut {NoStop}%
\bibitem [{\citenamefont {Huh}\ \emph {et~al.}(2010)\citenamefont {Huh},
  \citenamefont {Neff}, \citenamefont {Rauhut},\ and\ \citenamefont
  {Berger}}]{huh:2010anharm}%
  \BibitemOpen
  \bibfield  {author} {\bibinfo {author} {\bibfnamefont {J.}~\bibnamefont
  {Huh}}, \bibinfo {author} {\bibfnamefont {M.}~\bibnamefont {Neff}}, \bibinfo
  {author} {\bibfnamefont {G.}~\bibnamefont {Rauhut}}, \ and\ \bibinfo {author}
  {\bibfnamefont {R.}~\bibnamefont {Berger}},\ }\href@noop {} {\bibfield
  {journal} {\bibinfo  {journal} {Mol. Phys.}\ }\textbf {\bibinfo {volume}
  {108}},\ \bibinfo {pages} {409} (\bibinfo {year} {2010})}\BibitemShut
  {NoStop}%
\end{thebibliography}
\end{document}